\documentclass{article}
\usepackage{graphicx}
\usepackage{epsfig}
\usepackage{amsfonts}
\usepackage{slashed}
\usepackage{mathrsfs}
\usepackage{amssymb}
\usepackage{amsthm,amsmath,amssymb}
\usepackage{hyperref}
\usepackage{graphicx}       % Include figure files
\usepackage{dcolumn}        % Align table columns on decimal point
\usepackage{bm}             % bold math
\usepackage{amssymb}
\usepackage{amstext}
\usepackage{tensor}
\usepackage{cite}
\usepackage{subfig}
\usepackage{diagbox}
\usepackage[usenames]{color}
\usepackage{color}
\usepackage{graphics}
\usepackage{indentfirst}
\usepackage{booktabs}
\usepackage{float}
\usepackage{multirow}
\usepackage[section]{placeins}
\usepackage[capitalize]{cleveref}
\usepackage{array}
\newcolumntype{C}{>{8pt}c}

\usepackage{amssymb}

\date{\today}

\newcommand{\insertplot}[5]{\begin{figure}
		\hfill\hbox to 0.05in{\vbox to #5in{\vfill
				\inputplot{#1}{#4}{#5}}\hfill}
		\hfill\vspace{-.1in}
		\caption{#2}\label{#3}
\end{figure}}
\newcommand{\inputplot}[3]{
	\special{ps: plotfile #1}
}
\newcounter{fig}

\newcommand{\ee}{\end{equation}}
\newcommand{\eea}{\end{eqnarray}}
\newcommand{\be}{\begin{equation}}
\newcommand{\bea}{\begin{eqnarray}}

\begin{document}
	\title{\Large{\bf Excited Dirac stars with higher azimuthal harmonic index }}
	\vspace{1.5truecm}

	\author{
		{\large }%$^{\ddagger}$
		{\  Long-Xing Huang
  %\footnote{huanglx2023@lzu.edu.cn}
  },
		{\  Shi-Xian Sun
  %\footnote{120220908811@lzu.edu.cn}
  },
        {\  Rong Zhang
        %\footnote{zhangrong21@lzu.edu.cn}
        },
        {\  Chen Liang
        %\footnote{liangch2020@lzu.edu.cn}
        },
		and
		{\ Yong-Qiang Wang\footnote{yqwang@lzu.edu.cn, corresponding author
		}}
		\\
		\\
		$^{1}${\small Lanzhou Center for Theoretical Physics, 
		}
		\\
		{\small
			Key Laboratory of Theoretical Physics of Gansu Province,}
		\\
		{\small
			School of Physical Science and Technology,  }
		\\
		{\small
			Lanzhou University, Lanzhou 730000, China}
		\\
		$^{2}${\small Institute of Theoretical Physics $\&$ Research Center of Gravitation, 
		}
		\\
		{\small
			Lanzhou University, Lanzhou 730000, China}
		\\
		%$^{3}${\small Departamento de F\'isica da Universidade de Aveiro and CIDMA,
			%}
		%\\
		%{\small
			% Campus de Santiago, 3810-183 Aveiro, Portugal}
		%\\
		%$^{4}${\small
			%BLTP, JINR,
			%Joliot-Curie 6, Dubna 141980, Moscow Region, Russia}
	}
	\date{\today}
	
	\maketitle

	\begin{abstract}
	In this paper, we investigate the properties of the first excited state Dirac stars (DSs) with higher azimuthal harmonic index  (specifically, the azimuthal harmonic indexes $m_D$ = $3/2$, $5/2$, $7/2$), as well as the relationship between the ADM mass and angular momentum of Dirac stars with respect to frequency. Moreover, We find that the ergospheres of DSs appear at lower spinor field frequencies, and both the ergospheres and the distribution of the spinor field functions are asymmetric about the equatorial plane. Furthermore, we introduce the ground state scalar field and examine its impact on this system, which is known as the multi-state Dirac-boson stars (DBSs) model. We show various types of solution families for DBSs under both synchronized frequency $\omega$ and nonsynchronized frequencies and find that similar to DSs, the spinor field and the ergospheres of DBSs are also asymmetric about the equatorial plane, but the ergospheres appear at higher spinor field frequencies.

	\end{abstract}
	\newpage
	%%%%%%%%%%%%%%%%%%%%%%%%%%%%%%%%%%%%%%%%%%%%%%%%%%%%%%%%%%
	\section{INTRODUCTION}
	%%%%%%%%%%%%%%%%%%%%%%%%%%%%%%%%%%%%%%%%%%%%%%%%%%%%%%%%%%
	Since the famous Dirac equation was proposed by P. A. M. Dirac in 1928 \cite{Dirac:1928hu}, researchers have been exploring not only the wave function solutions that satisfy the Dirac equation but also the particle-like solutions (also referred to as solitons) that satisfy the Dirac equation. The initial research on this subject was conducted in flat spacetime. However, no such soliton structure exists in free fields in flat spacetime. Therefore, it is usually necessary to consider the spinor field with the self-interaction \cite{Ivanenko, Weyl, Heisenberg, Finkelstein:1951, Finkelstein:1956}. In 1970, considering the quartic self-interactions, M. Soler obtained a numerical solution of the lowest energy state nonlinear spinor field with a stable configuration in flat spacetime \cite{Soler:1970xp}. Subsequently, M. Soler again used numerical methods to study the extended charged Dirac particle model composed of the spinor field with quartic self-interactions and the Maxwell field \cite {Soler:1973rcj}. This result is the same as the one he and his collaborator Ranada calculated later using perturbation methods. \cite{Ranada:1973hna}. Furthermore, nonlinear spinor fields have also been used by A. F. Ranada and M. F. Ranada to explain strong interactions \cite{Ranada:1984tp}. In 1986, the stability and existence of nonlinear spinor fields were also numerically proven in Ref. \cite{Cazenave:1986pj}.
 
	However, if we consider the coupling of the spinor field to Einstein's gravity, the soliton solutions mentioned earlier still exist even without the self-interaction of the spinor field. These self-gravitating soliton solutions related to Einstein’s gravity have the desirable property of being free from singularities \cite{Leith:2020jqw} and are often referred to as ``Diarc Stars", ``Einstein-Dirac solitons" or ``Fermion stars" \cite{Leith:2021urf}. In 1957, D. R. Brill and J. A. Wheeler extended the Dirac equation in flat spacetime to the curved spacetime with the most general gravitational field while studying the interaction of gravity with neutrinos \cite{Brill:1957fx}. Later, T. D. Lee and Y. Pang used the Thomas-Fermi approximation in the curved spacetime to discover an approximate solution for a Dirac soliton, which they called ``Fermi soliton stars”. \cite{Lee:1986tr}. Furthermore, Ref. \cite{Talebaoui:1995xz} also formally discussed this direction.

	In addition to approximate solutions of the ED equation, how to solve the ED equation accurately is also a significant issue. People first study the exact numerical solution of Dirac stars in a static, spherically symmetric system. However, since a system consisting of only one spinor field does not possess a diagonal energy-momentum tensor, it is unable to form a spherically symmetric system. Therefore, the initial approach involved considering systems with two spinor fields (for more Dirac cases, see Refs. \cite{Leith:2020jqw, BakuczCanario:2020qmq}). Indeed, it is possible to employ a certain method to make all the off-diagonal elements of the total energy-momentum tensor of a system vanish when two spinor fields are considered together \cite{Herdeiro:2017fhv}. In 1999, F. Finster, J. Smoller, and S. T. Yau provided an exact numerical solution to the Einstein-Dirac equation for two spinor fields in spherically symmetric configurations. \cite{Finster:1998ws, Finster:1998ak}. Their work marked a significant breakthrough in this direction.

	Moreover, other matter fields can also be studied for their effect on the spinor field. Refs. \cite{Henriques1990, Lopes:1992np, DiGiovanni2020, DiGiovanni2022} discusses the fermion-boson stars using a method similar to ideal fluids. In 1999, F. Finster, J. Smoller, and S. T. Yau used numerical methods to study the charged spherically symmetric Dirac stars composed of two spinor fields and one Maxwell field \cite{Finster:1998ux}. And based on the study of the black hole solution of the Dirac-Einstein equation 
    \cite{Finster:1998ak, Ventrella:2003fu}, they further proved that the unique black hole solution of the static spherically symmetric Einstein-Dirac-Maxwell system is the Reissner-Nordström solution \cite{Finster:1998ju}. To better study the electroweak and strong interactions of Dirac particles, they considered the Einstein-Dirac-Yang-Mills model (EDYM) \cite{Finster:2000ps} composed of the Einstein field, the spinor field, and the Yang-Mills field in 2000 (Since the gauge field can transform under the rotation group representation of spinors, this paper can obtain a spherically symmetric static solution only through a spinor field and a Yang-Mills field).  Subsequently, they showed that for the spherically symmetric, static black hole solutions that satisfy EDYM equations, spinors would disappear outside the horizon by considering multi-particle systems and introducing angular momentum \cite{Finster:2000vy}. Ref. \cite{Liang:2022mjo} investigates the ground-state (All field functions have no nodes) Dirac-boson stars composed of a scalar field and a spinor field. For the synchronized frequency case (spinor field frequency $\omega_D$ equals scalar field frequency $\omega_S$) and nonsynchronized frequency case ($\omega_S\neq\omega_D$), respectively, several different types of solution families are given. Furthermore, the interaction between the Higgs field and the spinor field in the curved spacetime can be considered, which allows the mass of fermions to be generated through the Higgs mechanism \cite{Leith:2022jew}.

	Just like in flat spacetime, when considering the self-interaction (e.g., $(\bar{\Psi}\Psi)^n$) of the spinor field coupled with Einstein’s gravity, the physical properties of the system may also undergo significant changes \cite{Adanhounme:2012cm, Krechet:2014nda}. In recent years, people have begun to study the spherically symmetric Dirac stars with the self-interaction introduced by numerical calculations. In 2018, V. Dzhunushaliev and V. Folomeev analyzed the self-interaction effect on the Arnowitt-Deser-Misner mass of Dirac stars \cite{Dzhunushaliev:2018jhj}. Later, they continued to study the impact of Proca vector fields or Yang-Mills fields on DSs with the self-interaction \cite{Dzhunushaliev:2019kiy}. Then, C. A. R. Herdeiro and E. Radu compared the self-interacting Dirac stars, Proca stars, and boson stars and found that they share certain common properties \cite{Herdeiro:2020jzx}.

	What is also interesting is the study of the spinor field with nodes, which we call the excited states. Like the case of the ground state \cite{Daka:2019iix}, excited state Dirac stars also have stable solutions \cite{Finster:1998ws}. In Ref. \cite{Bohun:1999zdr}, C. S. Bohun and F. I. Cooperstock studied charged Dirac stars with the first and the second excited states. In recent years, research on the excited state Dirac stars has been extended to the third excited state DSs\cite{Herdeiro:2017fhv}, multi-fermion systems \cite{Leith:2021urf}, and multi-state Dirac stars \cite{Liang:2023ywv}.
 
	Although using axially symmetric geometry to study rotating Dirac stars can bring greater technical difficulties to numerical solutions, rotation is ubiquitous in all scales. So, studying rotating Dirac stars may lead to some interesting phenomena. Similar to the case of spherical symmetry. At first, people searched for approximate solutions of ED equations under axial symmetry configurations \cite{Mei:2011mh}. Later, the exact numerical solution of rotating Dirac stars was proposed  \cite{Herdeiro:2019mbz}. In 2021, C. A. R. Herdeiro et al. considered the effect of the $U(1)$ gauge field on rotating Dirac stars and found that the gyromagnetic ratio of Dirac stars is close to that of Kerr black holes \cite{Herdeiro:2021jgc}. Subsequently, Ref. \cite{Dzhunushaliev:2022wnd} introduced the effective state equation of spinor fluid to study rotating Dirac stars and proved that, in some specific cases, the main physical characteristics of Dirac stars could be close to those of typical rotating neutron stars. Other studies on the Einstein-Dirac equation include \cite{ Bronnikov:2004uu,Bronnikov:2009na,Fabbri:2011mg,Giulini:2012zu,Vicente:2018mxl,Ramazanoglu:2018hwk,Blazquez-Salcedo:2019qrz,Blazquez-Salcedo:2021udn,cui2022,Dzhunushaliev:2023sdq,Kain:2023ore}.

	In this paper, we initially studied the rotating DSs with the first excited state. Subsequently, we introduced the ground state scalar field and investigated its influence on the excited state DSs. Since the boundary conditions differ for spinor fields with angular quantum numbers $m_D=1/2$ and $m_D>1/2$, and Ref. \cite{Herdeiro:2019mbz} has already studied the case of rotating Dirac stars with $m_D = 1/2$, we therefore focused on excited-state rotating Dirac stars with the azimuthal indexes $m_D=3/2, 5/2, 7/2$.

	The paper is organized as follows. Sec. \ref{sec2} deals with the general framework of the solutions. In Sec. \ref{sec3}, the boundary conditions of the rotating DSs and DBSs are studied. In addition, we presented the numerical methods used to compute our model. We introduce some important physical quantities in Sec. \ref{sec4}. In Sec. \ref{sec5}, We show the numerical results of the equations of motion and exhibit the properties of the DSs and DBSs. We conclude in Sec. \ref{sec6} with a summary and outline for future work.

	%%%%%%%%%%%%%%%%%%%%%%%%%%%%%%%%%%%%%%%%%%%%%%%%%%%%%%%%%%
	\section{THE GENERAL FRAMEWOEK}
	\label{sec2}
	%%%%%%%%%%%%%%%%%%%%%%%%%%%%%%%%%%%%%%%%%%%%%%%%%%%%%%%%%%
    Firstly, we consider the minimal coupling of the spinor field to $3+1$-dimensional Einstein's gravity. Secondly, to investigate the influence of a scalar field on Dirac stars, we incorporate the scalar field into the action. Based on this, we can initially formulate the action for the most general Einstein-Dirac-Klein-Gordon model. We can set the scalar field to zero when we focus solely on Dirac stars. The specific form of the action is as follows:
    	\begin{equation}
			\mathcal{S}=\int d^4 x \sqrt{-g}\left[\frac{R}{16 \pi G}+\mathcal{L}_{S}+\mathcal{L}_{D}\right]. \label{eq:action}
		\end{equation}	
	Where $G$ represents the gravitational constant, $R$ is the Ricci scalar, $\mathcal{L}_{S}$ and $\mathcal{L}_{D}$ denote the Lagrangian densities of the scalar field and the spinor field, respectively. Specifically, they can be expressed as:
		\begin{align}      
			&\mathcal{L}_{S}=-g^{\alpha \beta} \partial_\alpha \Phi^* \partial_\beta \Phi-\mu_S^2 \Phi^* \Phi,  \\    
			&\mathcal{L}_{D}=-i\left[\frac{1}{2}(\hat{D}_\mu \bar{\Psi} \gamma^\mu \Psi-\bar{\Psi} \gamma^\mu \hat{D}_\mu \Psi)+\mu_D \bar{\Psi} \Psi\right].
			\label{eq:Lagrangian}
		\end{align}
	Here, $\Phi$ and $\Psi$ represent the complex scalar and spinor field, respectively. $\bar{\Psi}\equiv\Psi^\dagger\xi$ denotes the Dirac conjugate, where $\Psi^\dagger$ is the Hermitian conjugate of the spinor field. We can choose the gamma matrix $-i\gamma^0$ as the representation of the Hermitian matrix $\xi$ \cite{Dolan:2015eua}. $\mu_S$ and $\mu_D$ represent the scalar and spinor field masses, respectively. $\hat{D}_\mu=\partial_\mu+\Gamma_\mu$ represents the spin covariant derivative, where $\Gamma_\mu$ is the spin connection matrices. The gamma matrices $\gamma^\mu$ (using Greek letters $\mu$, $\nu$, $\ldots$ as spacetime indices) are the gamma matrices in the curved spacetime, which satisfy the following relationship with the gamma matrices $\gamma^a$ (using Latin letters $a$, $b$, $\ldots$ as spacetime indices) in flat spacetime,
	    \begin{equation}
			\gamma^\mu=e_a^\mu \gamma^a, \quad \gamma^a=e_\mu^a \gamma^\mu \label{eq:gammachange}.
		\end{equation}
    $e_a^\mu$ comprising an orthonormal basis is known as a tetrad or vielbein. In the $n$-dimensional spacetime, the tetrad can be written as a series of invertible $n\times n$ matrices. It satisfies the following relationship with the metric $g_{\mu\nu}$ in the curved spacetime:
    	\begin{equation}
	g_{\mu\nu}e^\mu_a e^\nu_b=\eta_{ab}.
		\end{equation}	
	Here, $\eta_{ab}=diag(-1,+1,+1,+1)$. In this paper, even though we are studying stationary spacetimes under the axially symmetric geometry, choosing the ``spheroidal-type" coordinates $(t, r, \theta, \phi)$ instead of cylindrical coordinates for writing the metric will be more convenient for numerical computations \cite{Herdeiro:2019mbz, Schunck:2003kk, Jetzer:1991jr}.
	  \begin{equation}
		d s^2=-e^{2 F_0} d t^2+e^{2 F_1}\left(d r^2+r^2 d \theta^2\right)+e^{2 F_2} r^2 \sin ^2 \theta\left(d \varphi-\frac{W}{r} d t\right)^2.
		\end{equation}

	The four metric functions  $F_0, F_1, F_2, W$ depend solely on the variables $r$ and $\theta$. The nonzero components of the vielbein $e^\mu_a$ corresponding to the metric can be written as:
	    \begin{align}
		&\quad \quad e_0^t=e^{-F_0(r,\theta)},\quad e_0^\varphi=\frac{e^{-F_0(r,\theta)W(r,\theta)}}{r},  \\
		&e_1^r=e^{-F_1(r,\theta)},\quad e_2^\theta=\frac{e^{-F_1(r,\theta)}}{r},\quad e_3^\varphi=\frac{e^{-F_2(r,\theta)}\csc(\theta)}{r}.   \label{eq:vierbein}        
		\end{align}
	Using the Weyl/chiral representation, we can express the gamma matrices in flat spacetime as follows \cite{Armendariz-Picon:2003wfx}:
	    \begin{equation}
		\gamma^0=\left(\begin{array}{cc}
			O & I \\
			I & O
		\end{array}\right), \quad {\gamma}^i=\left(\begin{array}{cc}
			O & \sigma_i \\
			-\sigma_i & O
		\end{array}\right), \quad i=1,2,3. \label{eq:gammamatrix}
		\end{equation}
	Where $I$ is the identity matrix and $\sigma_k$ represents the Pauli matrices:
	  \begin{equation}
		\sigma_1=\left(\begin{array}{ll}
			0 & 1 \\
			1 & 0
		\end{array}\right), \quad \sigma_2=\left(\begin{array}{cc}
			0 & -i \\
			i & 0
		\end{array}\right), \quad \sigma_3=\left(\begin{array}{cc}
			1 & 0 \\
			0 & -1
		\end{array}\right).
	\end{equation}
	Substituting Eqs. (\ref{eq:vierbein}) and (\ref{eq:gammamatrix}) into Eq. (\ref{eq:gammachange}), We have:
	   \begin{align}       
		&\quad \quad \quad \gamma^t=e^{-F_0(r,\theta)}\gamma^0,\quad \gamma^r=e^{-F_1(r,\theta)}\gamma^1   ,\\
		&\gamma^\theta=\frac{e^{-F_1(r,\theta)}}{r}\gamma^2,\quad \gamma^\theta=\frac{e^{-F_2(r,\theta)\gamma^3 \csc(\theta)+e^{-F_0(r,\theta)}\gamma^0 W(r,\theta)}}{r}.   
		\end{align}
	It can be proven that both sets of $\gamma$-matrices are representations of the Clifford algebra, and they satisfy the anticommutation relations.
	    \begin{equation}
		\left\{\gamma^\alpha, \gamma^\beta\right\}=2 g^{\alpha \beta}, \quad\left\{\gamma^a, \gamma^b\right\}=2     \eta^{a b}.
		\end{equation}
	Using the gamma matrices in the curved spacetime, we can obtain the spin connection matrices $\Gamma^\mu=\omega_{\mu ab}\gamma^a \gamma^b/4$, where $\omega_{\mu ab}$ is called the spin connection. The spin connection satisfies the following relationship:
	    \begin{equation}
		\omega_{\mu a b}=e_{a \nu} e_b^\lambda \Gamma_{\mu \lambda}^\nu-e_b^\lambda \partial_\mu e_{a \lambda}.
		\end{equation}
	Where $\Gamma^\lambda_{\nu\mu}$ is the Christoffel connection.
	
	Additionally, we choose the ansatz for the scalar field \cite{Kleihaus:2005me} and the spinor field \cite{Herdeiro:2019mbz} as follows:
	   \begin{equation}
		\Phi=\phi(r)e^{i(m_S \varphi-\omega_S t)},
		\end{equation}
		\begin{equation}
		\Psi_{(n)}=e^{i(m_D \varphi-\omega_D t)}\left(\begin{array}{c}
			A_{(n)}(r, \theta) + iB_{(n)}(r, \theta)\\
			C_{(n)}(r, \theta) + iD_{(n)}(r, \theta) \\
			-iA_{(n)}(r, \theta) - B_{(n)}(r, \theta) \\
			-iC_{(n)}(r, \theta) - D_{(n)}(r, \theta) 
		\end{array}\right), n=0,1,2,...
		\end{equation}
    $A_{(n)}$, $B_{(n)}$, $C_{(n)}$, and $D_{(n)}$ are real functions, and $m_S$ and $m_D$ are the azimuthal harmonic indexes. $m_S$ takes integer values, while $m_D$ takes half-integer values. The subscript $(n)$ for the spinor field indicates the number of nodes in the field function. It can be composed of radial node number $n_r$ and angular node number $n_\theta$, i.e., $n = n_r + n_\theta$. When $n = 0$, the resulting field function is referred to as the ground state, while for $n \geq 1$, it is referred to as an excited state. When not specifying the subscript, it refers to the ground or excited state of the DSs. $\omega_S$ and $\omega_D$ are the frequencies of the scalar and spinor fields, respectively, when $\omega_S = \omega_D = \omega$, the frequencies of the Dirac and scalar fields are referred to as synchronized frequencies, while when $\omega_S \neq \omega_D$, the case is referred to as nonsynchronized frequencies.
    
	By varying the action (\ref{eq:action}), we can obtain the Einstein field equation, the Dirac equation, and the Klein-Gordon equation:

	    \begin{equation}
		E_{\mu\nu} \equiv R_{\mu\nu}-\frac{1}{2} g_{\mu\nu}R-8\pi G((T_S)_{\mu\nu}+(T_D)_{\mu\nu})=0 ,\label{eq:EinsteinEq}
		\end{equation}
	    \begin{equation}
		\gamma^\mu\hat{D}_\mu\Psi-\mu_D \Psi=0,
		\end{equation}
		\begin{equation}
		\nabla^2\Phi-\mu_s^2\Phi=0.
		\end{equation}
	Where $R_{\mu\nu}$ is the Ricci curvature tensor, $(T_S)_{\mu\nu}$ and $(T_D)_{\mu\nu}$ are the energy-momentum tensors of the scalar field and the spinor field, respectively. Specifically, they are given by:
	    \begin{equation}
		(T_S)_{\mu\nu}=\partial_\mu\Phi^* \partial_{\nu}\Phi+\partial_\nu\Phi^* \partial_{\mu}\Phi-g_{\mu\nu}\left[\frac{1}{2}g^{\alpha \beta}(\partial_\alpha\Phi^* \partial_{\beta}\Phi+\partial_\beta\Phi^* \partial_{\alpha}\Phi)+\mu^2_S\Phi^*\Phi\right], \label{eq:emts}
	\end{equation}
	\begin{equation}
		(T_D)_{\mu\nu}=-\frac{i}{2}(\bar{\Psi}\gamma_\mu \hat{D}_\nu\Psi+\bar{\Psi}\gamma_\nu \hat{D}_\mu\Psi-\hat{D}_\mu\bar{\Psi}\gamma_\nu \Psi-\hat{D}_\nu\bar{\Psi}\gamma_\mu \Psi).       \label{eq:emtd}
	\end{equation}

	Finally, for the convenience of solving the Einstein equations, we utilize the following set of equations combination \cite{Herdeiro:2015gia}:
	    \begin{equation}
		\begin{aligned}
			& E_r^r+E_\theta^\theta-E_{\varphi}^{\varphi}-E_t^t=0, \\
			& E_r^r+E_\theta^\theta-E_{\varphi}^{\varphi}+E_t^t+2 W E_{\varphi}^t=0, \\
			& E_r^r+E_\theta^\theta+E_{\varphi}^{\varphi}-E_t^t-2 W E_{\varphi}^t=0, \\
			& E_{\varphi}^t=0 .
		\end{aligned}
	\end{equation}
	%	
		%%%%%%%%%%%%%%%%%%%%%%%%%%%%%%%%%%%%%%%%%%%%%%%%%%%%%%%%%%
	\section{BOUNDARY CONDITIONS AND NUMERICAL IMPLEMENTATION}
	\label{sec3}
		%%%%%%%%%%%%%%%%%%%%%%%%%%%%%%%%%%%%%%%%%%%%%%%%%%%%%%%%%%
		%\subsection{Boundary conditions}
		%%%%%%%%%%%%%%%%%%%%%%%%%%%%%%%%%%%%%%%%%%%%%%%%%%%%%%%%%%
	To solve the system of partial differential equations described in the Sec. \ref{sec2} , we utilize the assumption of regularity and asymptotic flatness, provide the conditions that the metric functions $F_i$ (where $i=0$, $1$, $2$) and the matter field functions $A_{(n)}$, $B_{(n)}$, $C_{(n)}$, $D_{(n)}$, $\phi$ need to satisfy at the boundaries.
	\begin{itemize}
			\item  The metric functions 
		\begin{equation}
			\begin{aligned}  
				&\partial_r F_i(r,0)=W(r,0)=0,\\
                &\partial_r F_i(r,\pi)=W(r,\pi)=0,\\   
                &\partial_r F_i(0,\theta)=W(0,\theta)=0,\\
                &\partial_r F_i(\infty,\theta)=W(\infty,\theta)=0. 
			\end{aligned}
		\end{equation}
		\item  The function of the scalar field 
  
		We require the scalar field to be zero at the boundary: 
		\begin{equation}
			\phi(0,\theta)=\phi(\infty,\theta)=\phi(r,0)=\phi(r,\pi)=0. 
		\end{equation}
		\item The functions of the spinor field  
  
		For the spinor field in the radial direction, we impose the following conditions:\\  
		 for $r=0$,
		\begin{equation}
			A_{(n)}(0,\theta)=B_{(n)}(0,\theta)=C_{(n)}(0,\theta)=D_{(n)}(0,\theta) =0,\
		\end{equation} 
		\end{itemize} 

		  \quad  and, for $r=\infty$,
  
		\begin{equation}
		 A_{(n)}(\infty,\theta)=B_{(n)}(\infty,\theta)=C_{(n)}(\infty,\theta)=D_{(n)}(\infty,\theta)=0.
		\end{equation}   

		However, for the angular direction, the boundary conditions differ significantly between the case of the higher azimuthal harmonic index we are studying ($m_D=3/2$, $5/2$, $7/2$) and the case of the lower azimuthal harmonic index ($m_D=1/2$). For the case of $m_D=1/2$, the boundary conditions are as follows:	
				\begin{equation}
			\partial_\theta A_{(n)}(r,0)=\partial_\theta B_{(n)}(r,0)=C_{(n)}(r,0)=D(r,0)=0,
		\end{equation}
					\begin{equation}
		A_{(n)}(r,\pi)= B_{(n)}(r,\pi)=\partial_\theta C_{(n)}(r,\pi)= \partial_\theta D_{(n)}(r,\pi)=0,
	\end{equation}
	
		But for $m_D>1/2$,we impose:
		\begin{equation}
			A_{(n)}(r,0)= B_{(n)}(r,0)=C_{(n)}(r,0)=D_{(n)}(r,0)=0,
		\end{equation}
			\begin{equation}
			A_{(n)}(r,\pi)= B_{(n)}(r,\pi)= C_{(n)}(r,\pi)=  D_{(n)}(r,\pi)=0.
	\end{equation}

	%%%%%%%%%%%%%%%%%%%%%%%%%%%%%%%%%%%%%%%%%%%%%%%%%%%%%%%%%%
	%\subsection{NUMERICAL IMPLEMENTATION}
	%%%%%%%%%%%%%%%%%%%%%%%%%%%%%%%%%%%%%%%%%%%%%%%%%%%%%%%%%%
	To facilitate numerical computations, we employ symmetry and perform the following scaling transformations on the equations \cite{Herdeiro:2019mbz}:
	    \begin{equation}
		\begin{aligned}
			&\widetilde{r}=\frac{ r}{\kappa},\quad \widetilde{\Phi}=\frac{\sqrt{2 \pi }}{M_{PI}}\Phi,\quad \widetilde{\omega}_D=\kappa\omega_D, \quad \widetilde{\mu}_S=\kappa \mu_S, \\
			&\widetilde{\Psi}=\frac{\sqrt{2 \pi }}{\sqrt{\mu}M_{PI}}\Psi,\quad \widetilde{\omega}_S=\kappa\omega_S,\quad  \widetilde{\mu}_D=\kappa \mu_D.
		\end{aligned}        
		\end{equation}
	Where $M_{PI}\equiv 1/\sqrt{G}$ represents the Planck mass, According to the action \ref{eq:action}, it can be deduced that $\kappa$ has dimensions of length. If we set $\kappa=\frac{1}{\mu_D}$, then:
	    \begin{equation}
		\widetilde{\mu}_D=1,\quad \widetilde{\mu}_S=\mu_S/\mu_D=l.
		\end{equation}
	Where $l$ is a constant, after the transformations mentioned above, the parameters of the problem under consideration are: ${\widetilde{\omega}_S, \widetilde{\omega}_D, m_S, m_D}$. In the following results, we set $m_S=1$ and consider $m_D=3/2$, $5/2$, $7/2$. Additionally, we introduce the following conformal transformation:
	    \begin{equation}
		\bar{x}=\frac{\widetilde{r}}{2+\widetilde{r}}.
		\end{equation}

	%%%%%%%%%%%%%%%%%%%%%%%%%%%%%%%%%%%%%%%%%%%%%%%%%%%%%%%%%%
	\section{QUANTITIES OF INTEREST}
		\label{sec4}		
	%%%%%%%%%%%%%%%%%%%%%%%%%%%%%%%%%%%%%%%%%%%%%%%%%%%%%%%%%%	
	The action of the matter fields are invariant under the $U(1)$ guaged transformation $\Phi \rightarrow e^{i\alpha}\Phi$, $\Psi \rightarrow e^{i\beta}\Psi$ with constants $\alpha$ and $\beta$ \cite{Liebling:2012fv, Daka:2019iix}. According to Noether's theorem, This implies the existence of conserved 4-currents corresponding to these two matter fields \cite{Kumar:2014kna, Blazquez-Salcedo:2019qrz},
    \begin{equation}
	j_S^\mu=-i\left(\psi^* \partial^\mu \psi-\psi \partial^\mu \psi^*\right), \quad j_D^\mu=\bar{\Psi} \gamma^\mu \Psi.
	\end{equation}
	They satisfy the following conditions:
	    \begin{equation}
		j^\mu_{S; \, \mu}=0, \quad j^\mu_{D ; \, \mu}=0.
		\end{equation}
	Substituting the above ansatz into this expression, we obtain the time component of the $4$-currents as: 
	    \begin{equation}
		j_S^t =2e^{-2F_0}(\omega_S-\frac{m_SW}{r})\phi^2,\quad
		j_D^t =2e^{-2F_0}(A^2+B^2+C^2+D^2).
		\end{equation}
	Furthermore, the $\varphi-t$ component of the energy-momentum tensor is given by:
	    \begin{equation}
		(T_S)^t_\varphi=2e^{-2F_0}m_S(\omega_S-\frac{m_SW}{r})\phi^2,
	\end{equation}
	
	\begin{equation}
		\begin{aligned}
			(T_D)^t_\varphi=&e^{-F_0} m_D\left(A_{(n)}^2+B_{(n)}^2+C_{(n)}^2+D_{(n)}^2\right)\\
			&+e^{-F_0-F_1+F_2}\sin \theta \left(A_{(n)} C_{(n)}+B_{(n)} D_{(n)})\left[1+r\left(F_{2, r}-F_{0, r}\right)\right]\right. \\
			&\left.+\frac{1}{2}e^{-F_0-F_1+F_2}\sin \theta\left(A_{(n)}^2+B_{(n)}^2-C_{(n)}^2-D_{(n)}^2\right)\left(F_{0, \theta}-\cot \theta-F_{2, \theta}\right)\right.\\
			&\left.+2 e^{2F_0+F_2} r \sin \theta  \left(\omega_D-\frac{m_D W}{r}\right)(B_{(n)} C_{(n)}-A_{(n)} D_{(n)})\right.
		\end{aligned}
	\end{equation}
	By integrating the timelike components of these two four-currents $(j_K)^t$ and the $(T_K)^t_\varphi$ over a spacelike hypersurface $\mathscr{S}$, we obtain the corresponding Noether charge $Q_K$ and angular momentum $J_K$ (where $K$ represents $S$ or $D$):
	    \begin{equation}
		Q_K^\mu=\int_\mathscr{S} j^t_K, \quad J_K^\mu=\int_\mathscr{S} (T_K)^t_\varphi.
		\end{equation}
	the total Noether charge $Q=Q_S+Q_D$, and the total angular momentum $J^\mu=J^\mu_S+J^\mu_D$.
	At the microscopic level, this Noether charge $Q$ is equivalent to the particle number $N$\cite{Hartmann:2010pm, Yoshida:1997qf,Collodel:2017biu,Bernal:2009zy,Bezares:2017mzk}.
 
	The ADM mass of $n=3+1$ dimensional DBSs can be obtained from the integral of the Komar energy density:
            \begin{equation}
    	M=2 \pi \int dr d\theta r^2e^{F_0+2F_1+F_2} ((T_S)^{\mu}_{\;\mu}+(T_D)^{\mu}_{\;\mu}-2(T_S)^{t}_{\;t}-2(T_D)^{t}_{\;t}).
    \end{equation} 
	On the other hand, the mass $M$ of DBSs and the total angular momentum $J$ can also be obtained from the asymptotic behavior of the metric functions at infinity,
	    \begin{equation}
		\begin{aligned}
			&g_{tt}|_{r=\infty}=-e^{2F_0}+e^{2F_2}W^2\sin^2{\theta}=-1+\frac{2GM}{r}, \\
			&g_{\varphi t}|_{r=\infty}=-e^{2F_0}W r\sin^2{\theta}=-\frac{2GJ}{r}\sin^2{\theta}.
		\end{aligned}
		\end{equation}
	%%%%%%%%%%%%%%%%%%%%%%%%%%%%%%%%%%%%%%%%%%%%%%%%%%%%%%%%%%
	\section{NUMERICAL RESULTS}
	\label{sec5}		
	%%%%%%%%%%%%%%%%%%%%%%%%%%%%%%%%%%%%%%%%%%%%%%%%%%%%%%%%%%	
	Our study primarily focuses on two aspects. On the one hand, we study the properties of the first excited state of Dirac stars and analyze their ergospheres. On the other hand, we utilize the Dirac-boson stars model to investigate the influence of the ground state scalar field on Dirac stars.
 
    The definition of the excited state is similar to atomic theory and quantum mechanics. For example, the first excited state of a hydrogen atom has an electron in the $2s$ or $2p$ orbital, corresponding to the radial and angular nodes of the hydrogen atom wave function, respectively. Therefore, we refer to the spinor field with one radial node ($n_r=1$) as the $D_1$ state; the ground state scalar field and the ground state spinor field without nodes are denoted as $S_0$, and $D_0$ states, respectively. Finally, the DBSs composed of the ground state scalar field and the excited state spinor field are represented as $D_1S_0$. In this paper, the four spinor field functions that constitute the first excited state of DSs each have one node in the radial direction, and there are no nodes in the angular direction (i.e., $n=n_r=1$, $n_\theta=0$). As no solutions with a lower total number of nodes have been found at present, we refer to this type of solution as the first excited state of DSs. 	
	
	Our solving process is based on the finite element methods, where we solve the weak form of the equations. The iteration process follows the Newton-Raphson method. The grids we use have sizes of $N_r \times N_\theta$ in the integration domain $x\in[0,1)$ and $\theta\in[0,\pi]$. Typically, we set $N_r=240$ and $N_\theta=120$. The relative error of our numerical solutions is typically below $10^{-4}$.
	%%%%%%%%%%%%%%%%%%%%%%%%%%%%%%%%%%%%%%%%%%%%%%%%%%%%%%%%%%
	\subsection{$D_1$ state Dirac stars}
\begin{figure}[!htbp]
		\centering
		\subfloat{  % []内可单独为每个小图命名。默认按照(a)(b)...的顺序命名，若省去[]则小图不命名。
			\includegraphics[width=7cm]{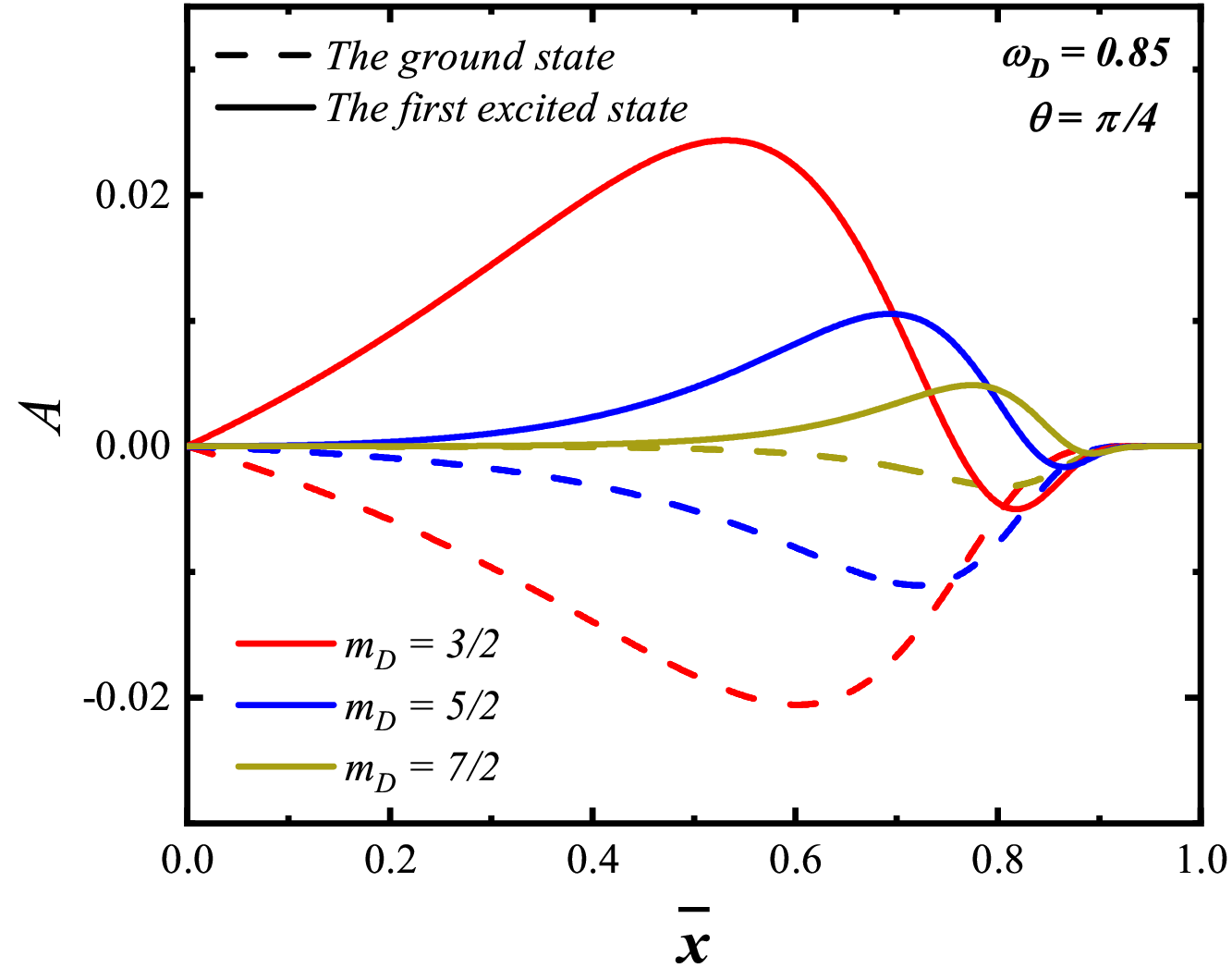}}
		\subfloat{
			\includegraphics[width=7cm]{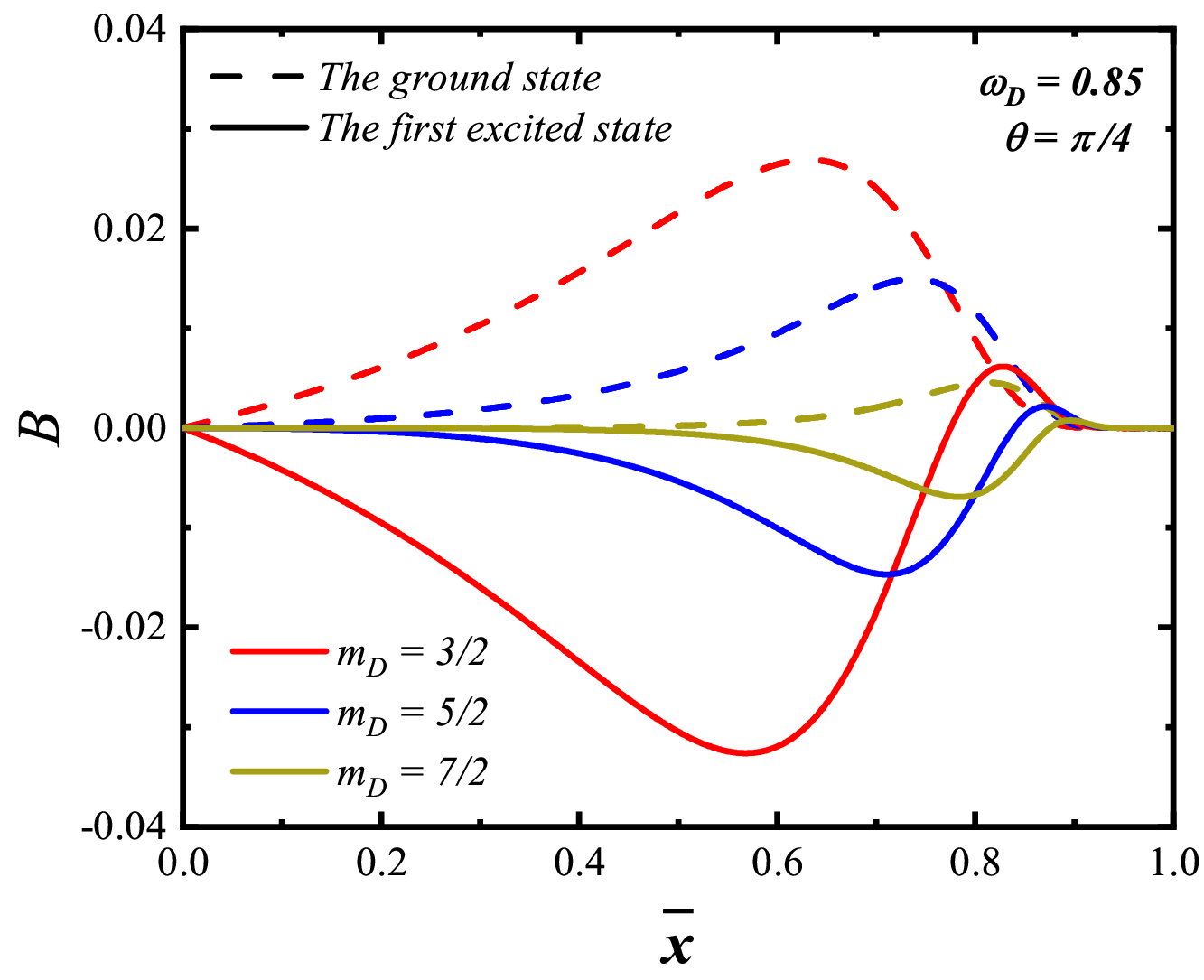}}
		\quad
		\subfloat{
			\includegraphics[width=7cm]{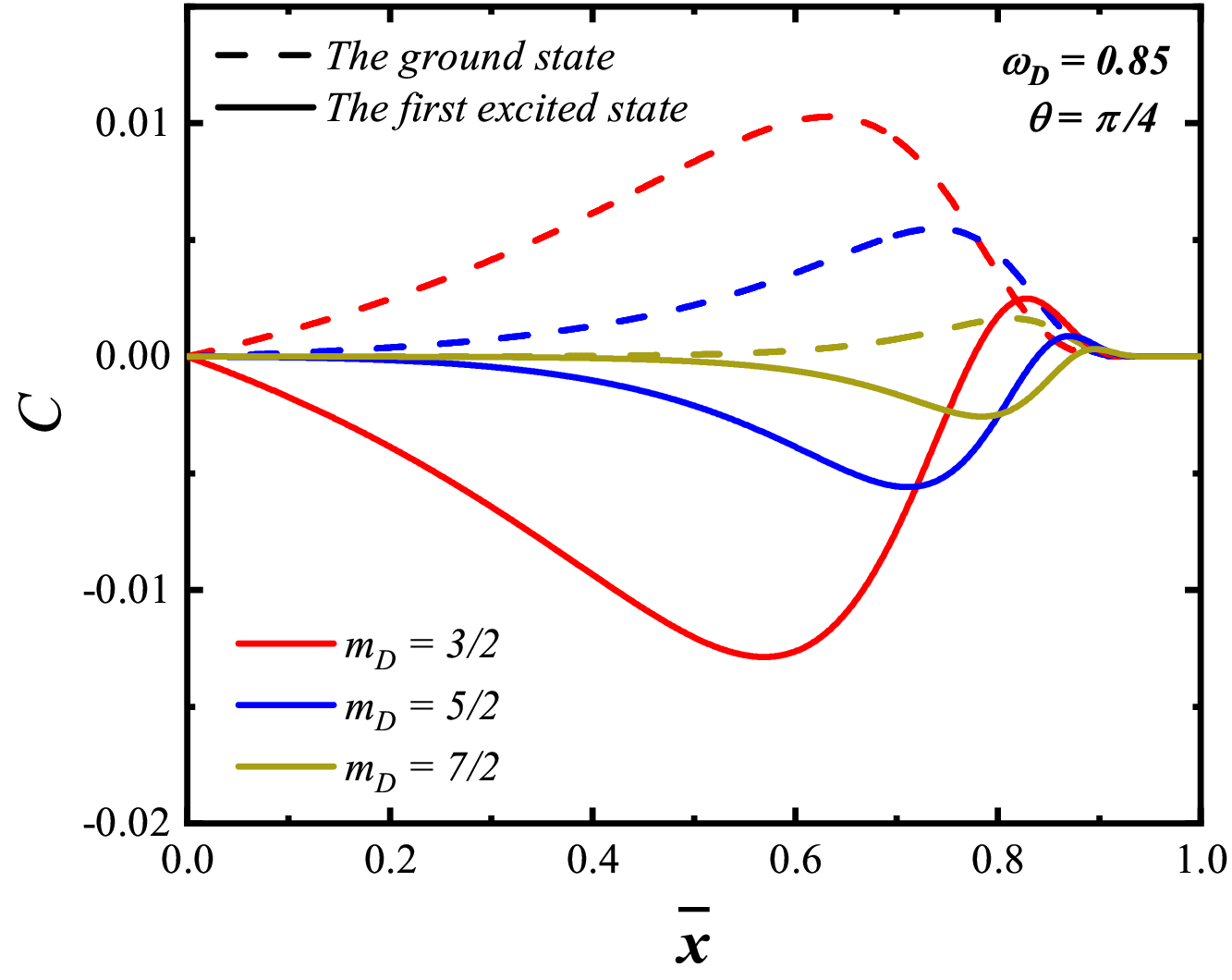}}
		\subfloat{
			\includegraphics[width=7cm]{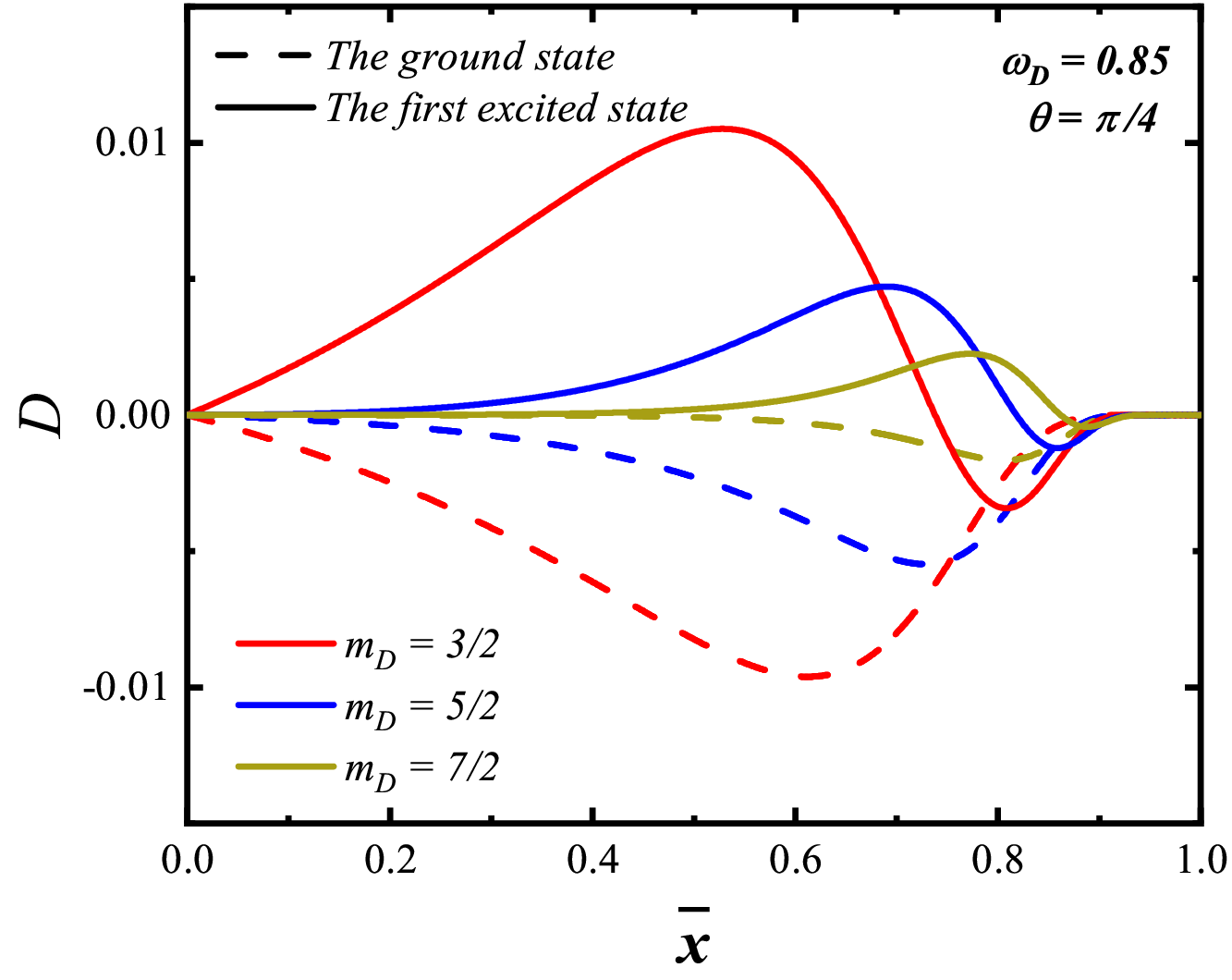}}
   		\quad
   		\subfloat{
				\includegraphics[width=7cm]{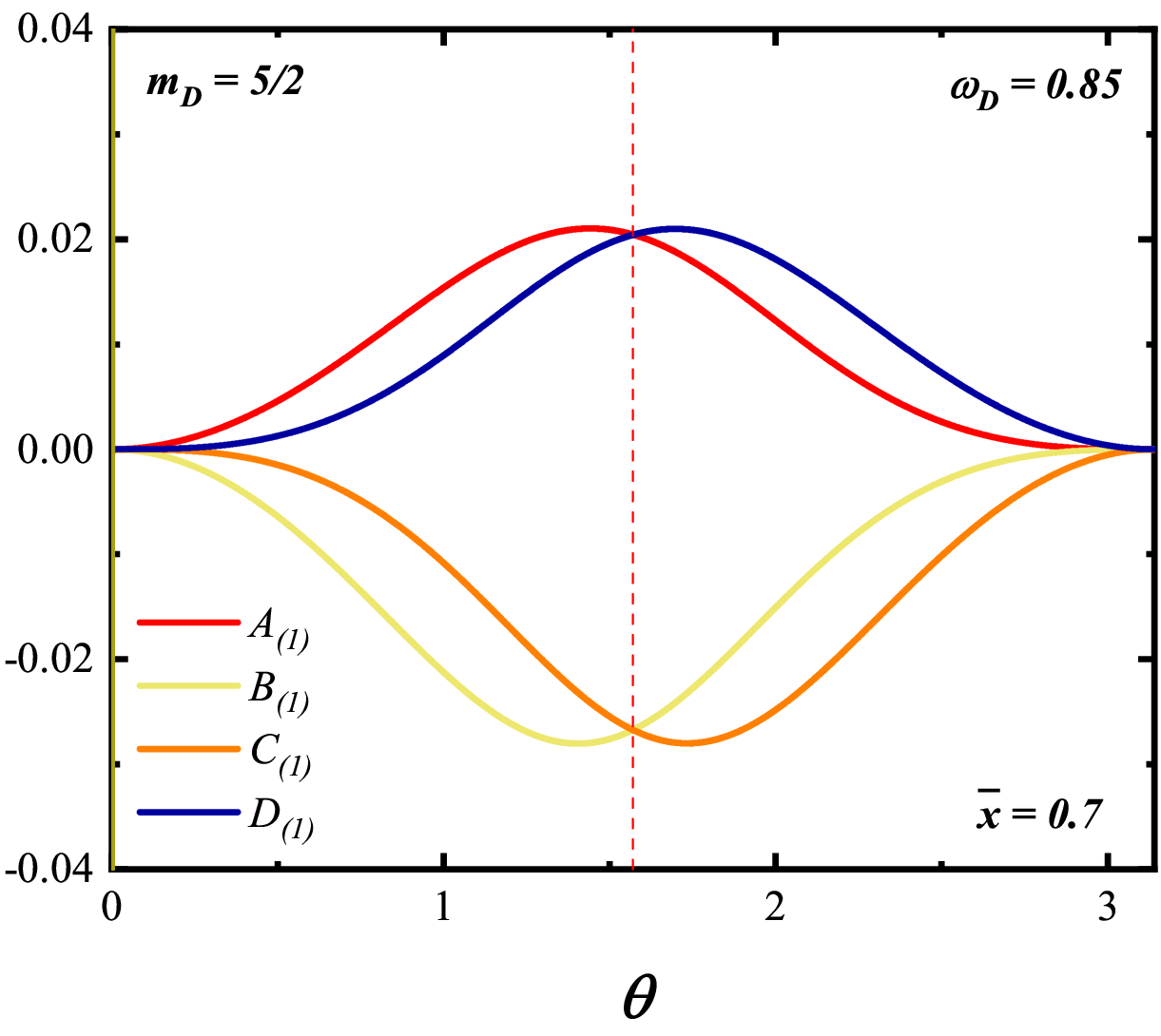}
				\label{fig: Dergospheresingled}  
			}
		\caption{ Top and middle panels: The distribution of the ground state (dash line) and the first excited state (solid line) spinor field as functions of $\bar{x}$ on the cross-section $\theta=\pi/4$, with the azimuthal indexes being $m_D=3/2$ (red line), $m_D=5/2$ (blue line), and $m_D=7/2$ (yellow line). Bottom panel: The field distribution of the spinor field with the azimuthal indexes being $m_D=5/2$ as a function of $\theta$ on the $\bar{x}=0.7$. The red dashed line represents the intersection of the equatorial plane with $\theta=\pi/2$. All solutions have the same frequency, $\omega_D=0.85$}
		\label{fig:FunctionOf0grand1st}
	\end{figure}

	This section will discuss the first excited state DSs ($D_1$) with one radial node and compare its physical quantities with the ground state DSs ($D_0$) with zero nodes. In the top and bottom panels of Fig. \ref{fig:FunctionOf0grand1st}, we present the distribution functions of the spinor field field $A$, $B$, $C$, and $D$ for the example of $\theta=\pi/4$ with frequencies $\omega_D=0.85$ and the azimuthal harmonic indexes $m_D=3/2$, $m_D=5/2$, and $m_D=7/2$. The dashed lines represent the ground state, while the solid lines represent the first excited state. These curves show that as the azimuthal harmonic indexes increase, both for the excited state and the ground state, the values of $A^{max}_{abs}$, $B^{max}_{abs}$, $C^{max}_{abs}$, and $D^{max}_{abs}$ (the maximum absolute values of the four field functions) decrease. Additionally, in the bottom panel of Fig. \ref{fig:FunctionOf0grand1st}, we provide an example for $m_D=5/2$, showing the graphs of four field functions at $\bar{x}=0.7$. It can be observed that all four field functions exhibit asymmetry with respect to the equatorial plane (indicated by the red dashed lines in the figure). 

\begin{figure}[!htbp]
	\centering
	\subfloat[]{  % []内可单独为每个小图命名。默认按照(a)(b)...的顺序命名，若省去[]则小图不命名。
		\includegraphics[width=7cm]{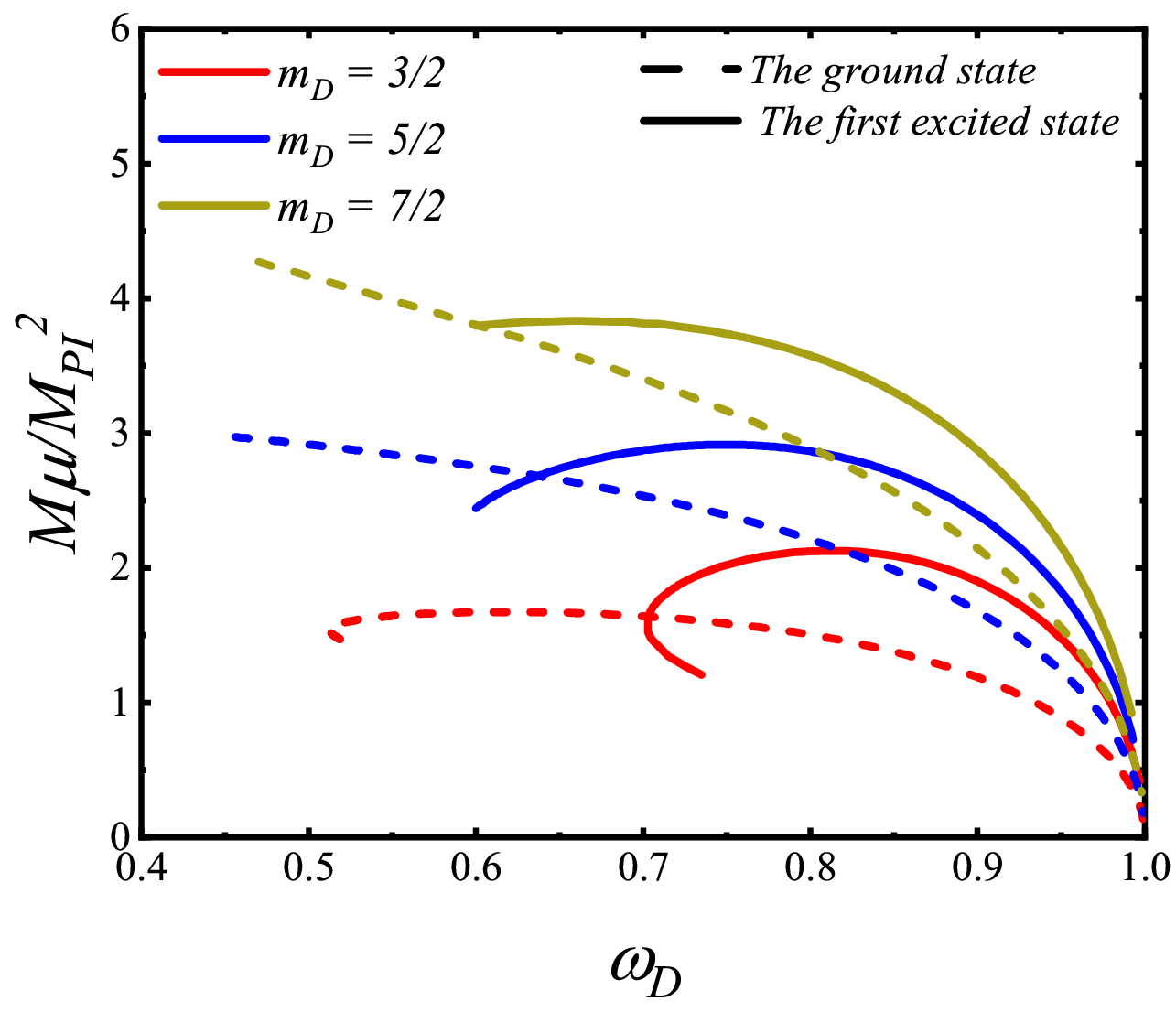}
		\label{fig: FunctionOf0grandMJN_a}	
		}
	\subfloat[]{
		\includegraphics[width=7cm]{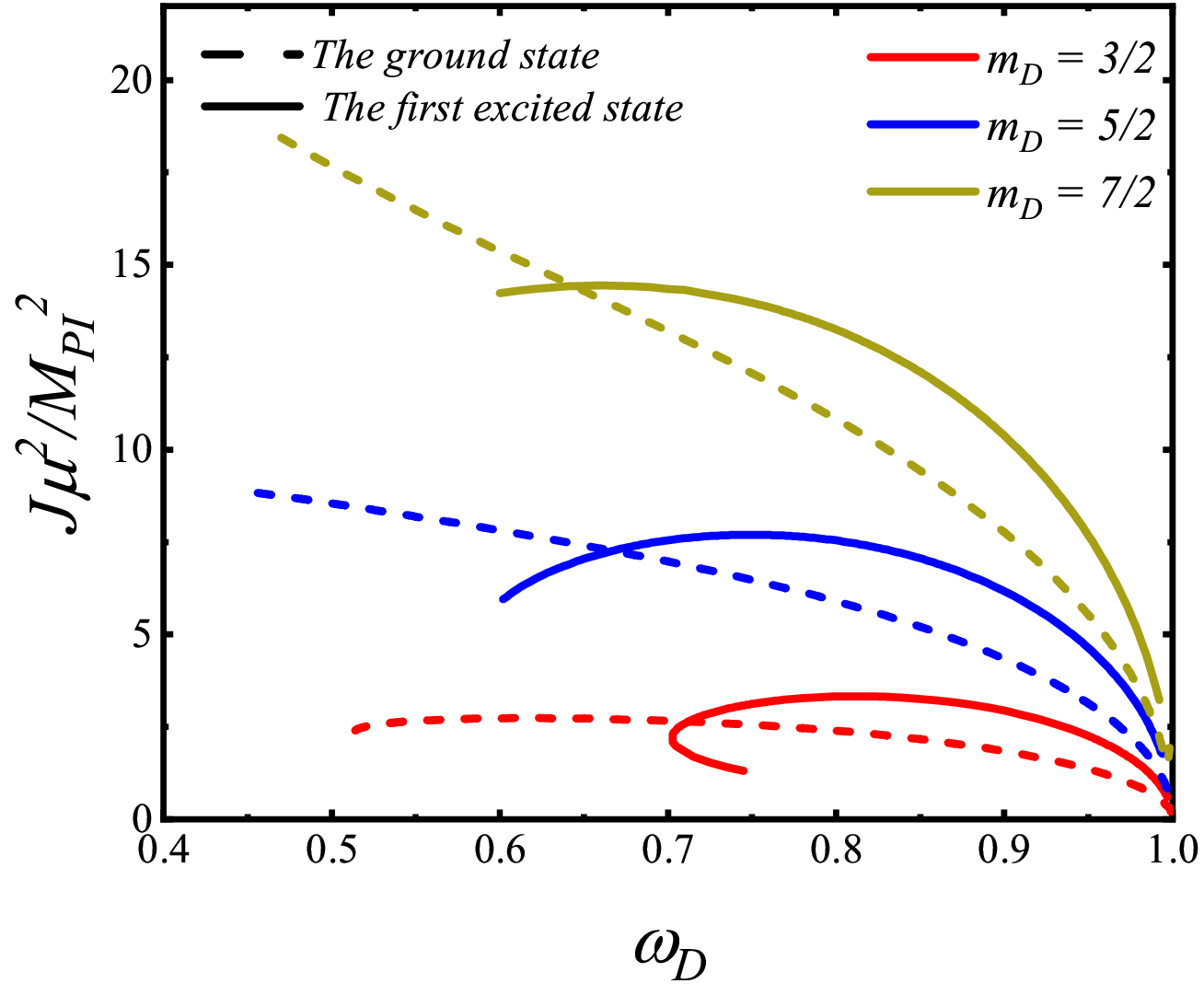}
		\label{fig: FunctionOf0grandMJN_b}	
		}
	\quad
	\subfloat[]{
		\includegraphics[width=7cm]{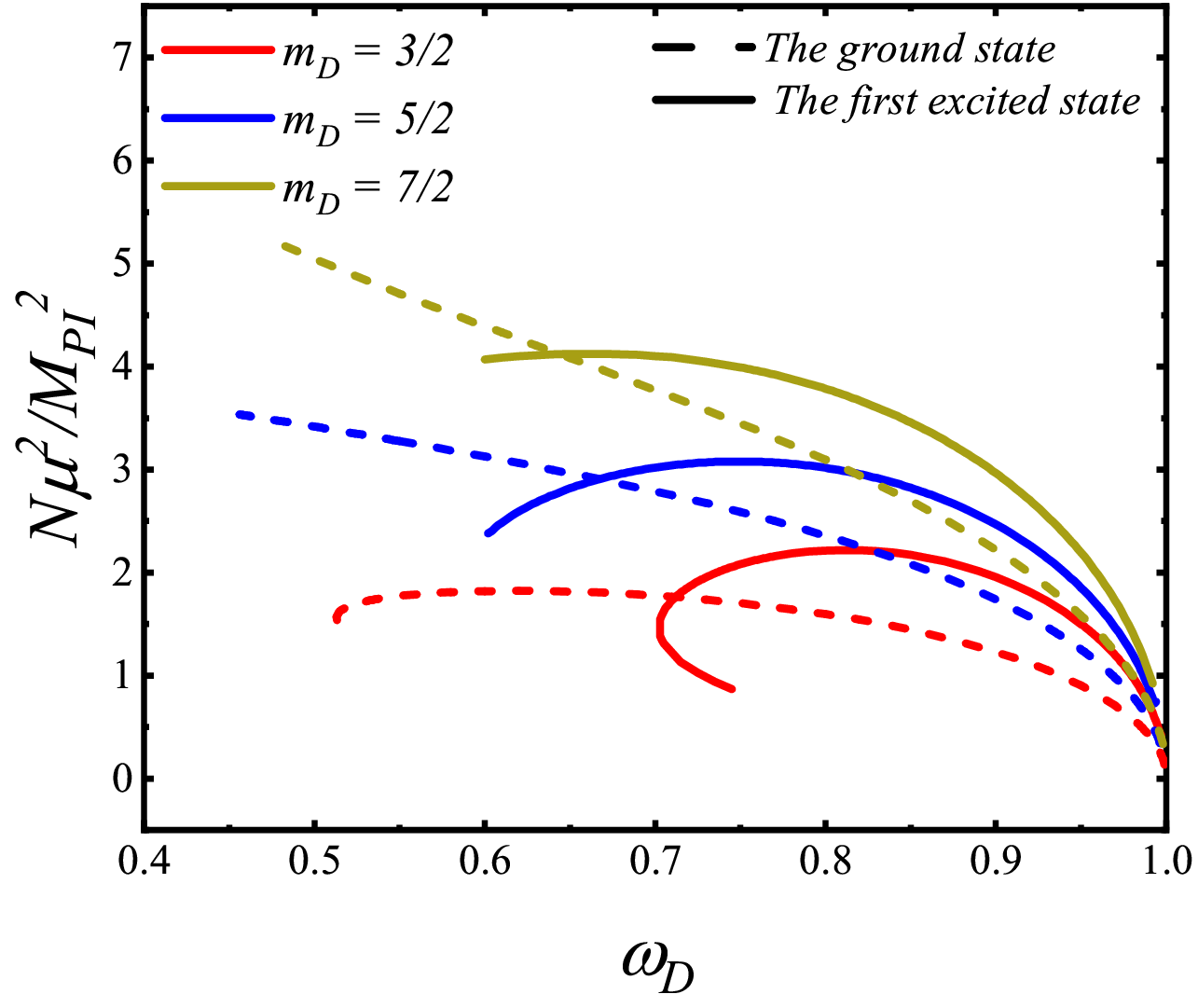}
		\label{fig: FunctionOf0grandMJN_c}
		}
	\subfloat[]{
		\includegraphics[width=7cm]{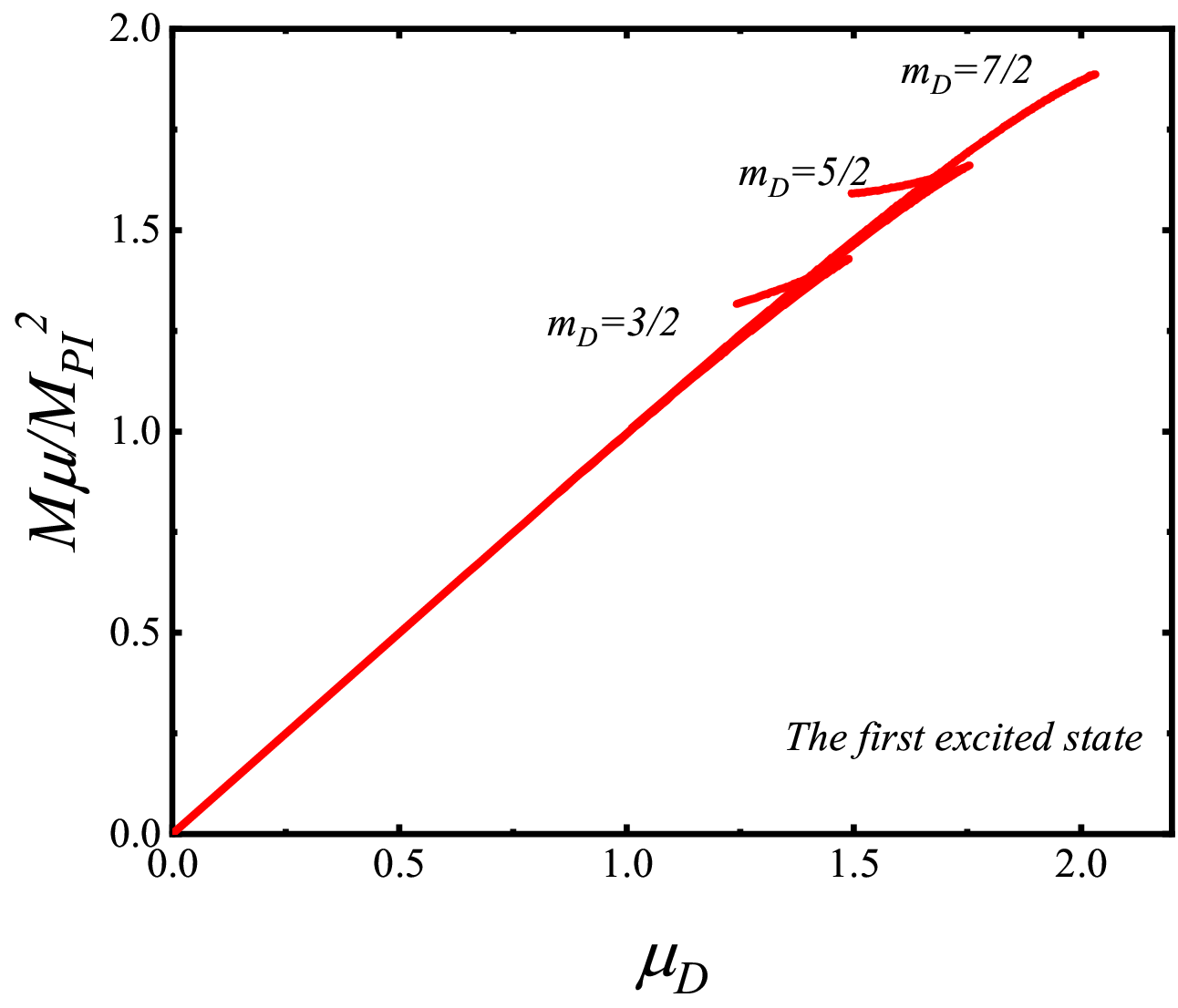}
		\label{fig: FunctionOf0grandMJN_d}
		}
	\caption{Top and bottom left panel: The ADM mass $M$ (top left panel), the angular momentum $J$ (top right panel), and the particle numbers $N$ (left bottom panel) of the ground state (dash line) and the first excited (solid line) as a function of the frequency $\omega_D$ with several values of the azimuthal index. Bottom right panel: When the particle number $N=1$, for the azimuthal index $m_D=3/2$, $5/2$, $7/2$, the relationship of the first excited spinor field mass $\mu_D$ between ADM mass $M$.}
	\label{fig: FunctionOf0grandMJN}
	\end{figure}

	To further study the physical properties of the obtained DSs, in Fig. \ref{fig: FunctionOf0grandMJN}, we exhibit the relationship between the ADM mass $M$ (Fig. \ref{fig: FunctionOf0grandMJN_a}), the angular momentum $J$ (Fig. \ref{fig: FunctionOf0grandMJN_b}), and the particle number $N$ (Fig. \ref{fig: FunctionOf0grandMJN_c}) and spinor field frequency $\omega_D$ for the ground state and first excited state DSs with three azimuthal harmonic indices. The dashed lines represent the $D_0$ state (the ground state), while the solid lines represent the $D_1$ state (the first excited state). The red, blue, and yellow lines represent the three azimuthal harmonic indices. The $M$, $J$, and $N$ vs. $\omega_D$ curves for the Dirac stars initially start from $\omega_D=1$ (vacuum solution) and reach a minimum frequency (the solution of the first branch). In the case of $m_D=3/2$, after getting the minimum frequency, the mass continues to decrease further as the frequency increases until it reaches the sub-maximum value of the frequency (the solution of the second branch). For $m_D=5/2$ and $m_D=7/2$, we speculate that the second branch might also exist, but we have not found it yet due to computational precision and errors. 
 
	For the three different angular quantum numbers, within a certain frequency range, the $D_1$ excited state exhibits greater mass, angular momentum, and particle number compared to the $D_0$ ground state. Additionally, it can be observed that DSs have a maximum ADM mass, angular momentum, and particle number, with this maximum value increasing with the azimuthal harmonic index. These parameters are provided in Table \ref{tab: SingleDirac}, and it can be seen from the table that, for a given angular quantum number, the frequencies $\omega_D$ (as indicated in the second row of the table) corresponding to the maximum mass $M^{max}$, the angular momentum $J^{max}$, and the particle number $N^{max}$ are the same. For $m_D=3/2$, $\omega_D=0.812$, for $m_D=5/2$, $\omega_D=0.750$, and, for $m_D=7/2$, $\omega_D=0.660$.
 
	Since we consider fermions that satisfy the Pauli exclusion principle, in Fig. \ref{fig: FunctionOf0grandMJN_d}, we go beyond the analysis of classical field theory and impose the condition of particle number $N=1$ for the excited state spinor field, showing the relationship between the mass $M$ and the field mass $\mu_D$. It can be observed that both the ADM mass and $\mu_D$ are bounded, and their maximum values increase with the azimuthal harmonic index.  Within a certain range, there is a linear relationship between $M$ and $\mu_D$. These three jagged curves are similar to the cases of the spherically excited state Dirac stars \cite{Herdeiro:2019mbz}, the lower azimuthal harmonic index rotating Dirac stars \cite{Herdeiro:2017fhv}, and the gauged Dirac stars\cite{Herdeiro:2021jgc}.
	\begin{table}[!htbp]
		\centering
		\begin{tabular}{|l||c|c|c|c|}
			\hline   & $\omega_D\left (M^{\max  },J^{\max  },Q^{\max  }\right)$ & $M^{\max }$ & $J^{\max }$ & $N^{\max }$  \\
			\hline $m_D=3/2$ & 0.812 & 2.128 & 3.327  & 2.218  \\
			$m_D=5/2$ & 0.750 & 2.914 & 7.697 & 3.079  \\
			$m_D=7/2$ & 0.660 & 3.832 & 14.437 & 4.125 \\
			\hline
		\end{tabular}
		\caption{The maximum mass, angular momentum, and particle number ($M^{\max }, J^{\max }, Q^{\max }$) of DSs with $m_D=3/2$ (top), $m_D=5/2$ (middle), $m_D=7/2$ (bottom) and their corresponding frequencies $\omega_D$.}
		\label{tab: SingleDirac}
	\end{table}

	At the end of this section, we will analyze the ergospheres of the first excited DSs. It should be noted that, for the convenience of plotting and comparing the results, we use Cartesian coordinates $\rho$, $y$, $z$ ($\rho=(2x\sin{\theta}\cos{\varphi})/(x-1)$, $\rho=(2x\sin{\theta}\sin{\varphi})/(x-1)$, $z=(2x\cos{\theta})/(x-1)$) instead of the previously transformed spherical coordinates for the diagrams of ergospheres. 

\begin{figure}[!htbp]
		\centering		
		\subfloat{  % []内可单独为每个小图命名。默认按照(a)(b)...的顺序命名，若省去[]则小图不命名。
			\qquad
			\qquad
			\includegraphics[width=6cm]{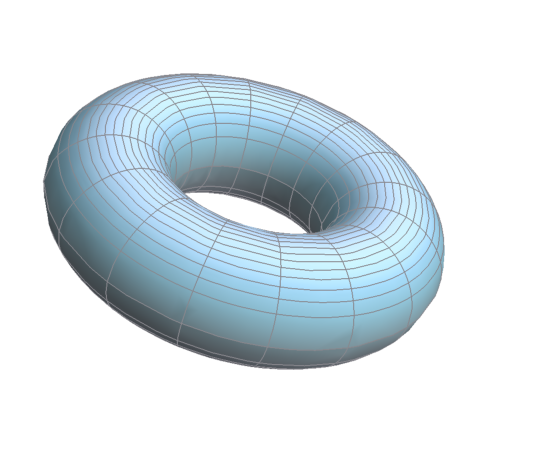}
			\label{fig:ergospheresinglea}
		}
		\subfloat{
			\includegraphics[width=6cm]{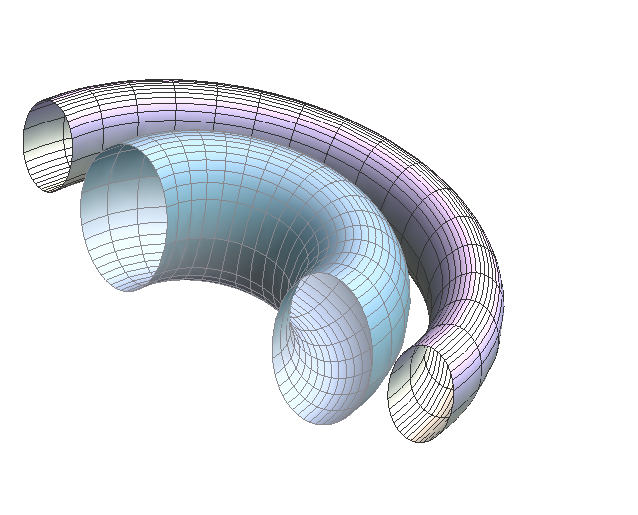}
			\label{fig:ergospheresingleb}
		}
		\quad
  		\subfloat{
			\includegraphics[width=6cm]{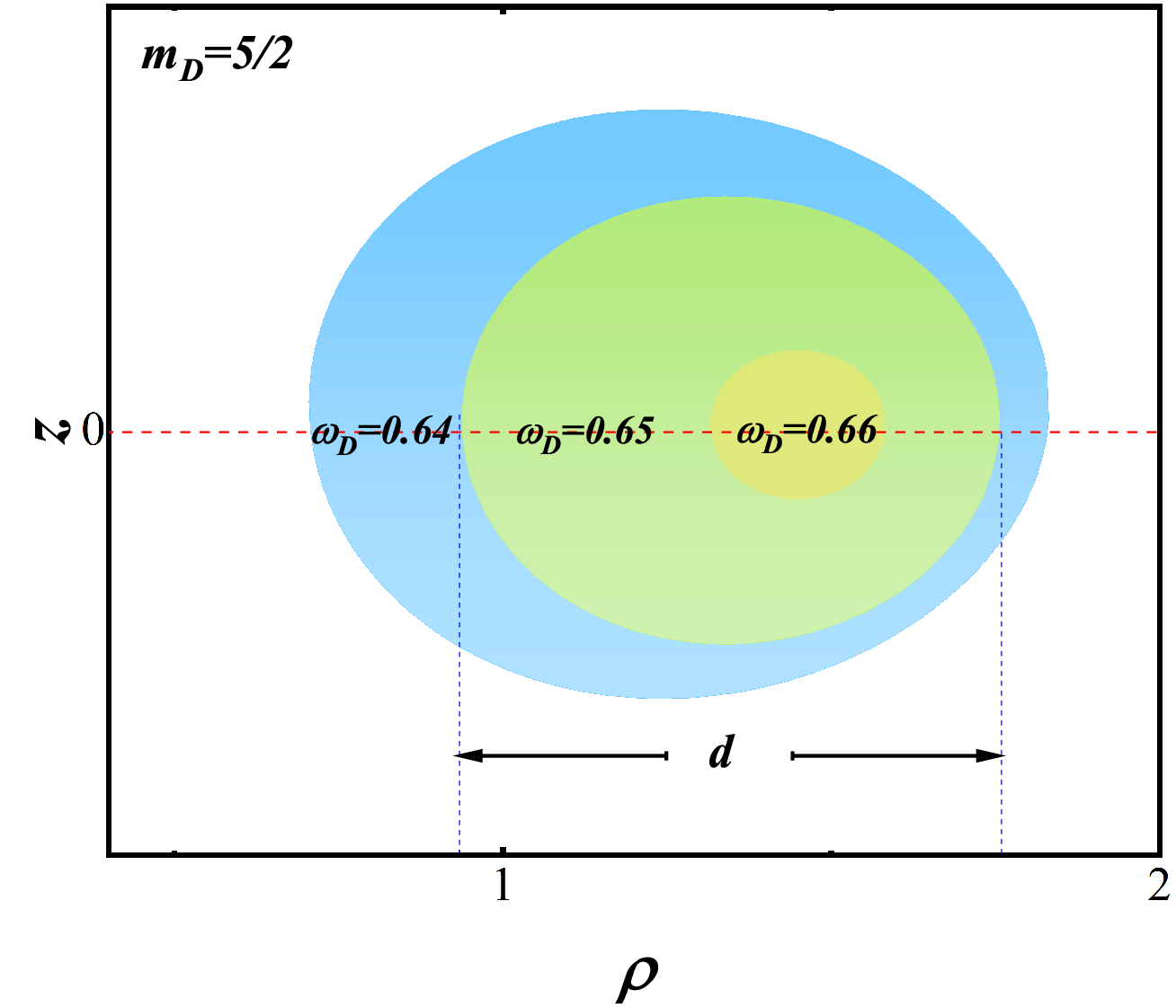}
			\label{fig:ergospheresinglec}  
		}
		\subfloat{
			\includegraphics[width=6.4cm]{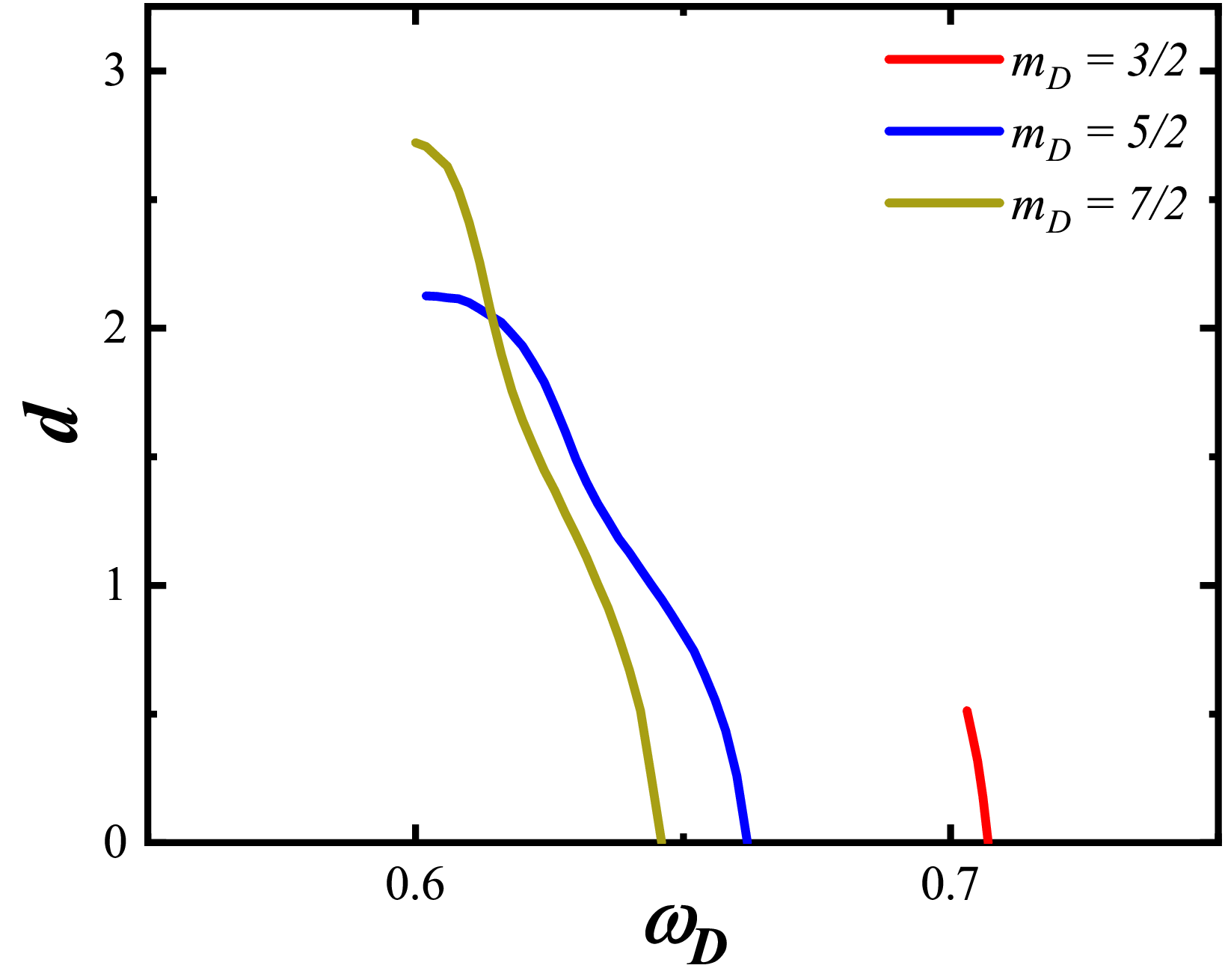}
			\label{fig:ergospheresingled}  
   		}
   		\quad
			\subfloat{
				\includegraphics[width=6cm]{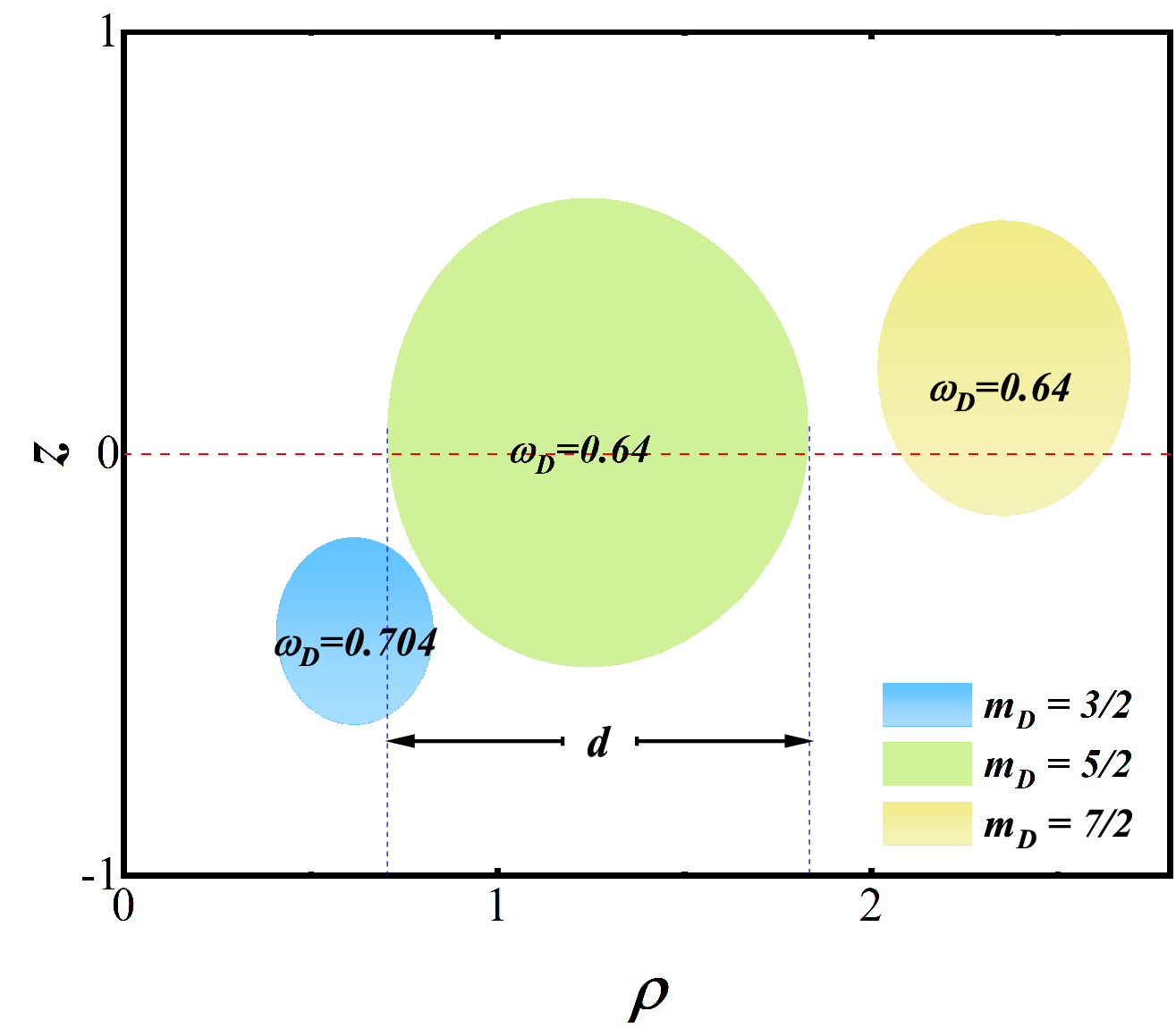}
				\label{fig: Dergospheresinglee}
			}  
		 
		\caption{ Top panel: The left panel shows the distribution of the ergosurface in the domain of existence for the first excited state DSs with $m_D=5/2$, $\omega_D=0.64$, and in the right panel, the two semi-circular structures represent the half of the ergosurface distribution for $m_D=5/2$(blue), $7/2$(purple) with $\omega_D=0.64$. Middle left panel: When $m_D=5/2$, the cross-sectional views of the ergosurface of the first excited state DSs with $\omega_D=0.64, 0.65, 0.66 $ on the $\rho\text{-}z$ plane. The red dashed line is the intersection of the equatorial plane with the $\rho\text{-}z$ plane. Middle right panel: The variation of the width $d$ of the ergospheres for the different azimuthal harmonic index of the first excited state as a function of frequency $\omega_D$. The definition of the width $d$ of the ergospheres is shown in this figure. Bottom Panel: The cross-sectional plots of the ergospheres of the first excited state DSs on the $\rho\text{-}z$ plane, where blue, green, and yellow represent azimuthal harmonic indexes of $m_D=3/2$ ($\omega_D=0.704$), $m_D=5/2$ ($\omega_D=0.64$), and $m_D=7/2$ ($\omega_D=0.64$), respectively. According to the symmetry of the ergospheres, we only present the right half of the cross-sections of the ergospheres (i.e., $\rho>0$).}
		\label{fig:ergospheresingle}		
		\end{figure}

	In the top panel of Fig. \ref{fig:ergospheresingle}, we provide a 3-dimensional plot of the ergosurface distribution of the first branch of excited DSs as an example with $\omega_D=0.64$, $m_D=5/2$ (blue the left panel and the interior of the right panel), and $m_D=7/2$ (purple the external of the right panel). The left plot represents only the ergosurface for $m_D=5/2$, while the right plot depicts half of the ergosurfaces for both azimuthal harmonic indexes of DSs, taking symmetry into account.  The blue or purple surfaces satisfy $g_{tt}=0$, and the interior of these surfaces represents the region where the ergospheres exist. It can be observed that the ergospheres exhibit a toroidal structure.

	To study how the ergospheres change with frequency $\omega_D$, considering symmetry, we present cross-sectional plots of the ergospheres of DSs on the $\rho\text{-}z$ plane ($\rho>0$) for three different frequencies, using $m_D=5/2$ as an example in the left of the middle panel. On the one hand, it can be seen that, as the frequency decreases, the center of the cross-section of the ergospheres continuously moves toward the $z$-axis. If we define the difference between the maximum and minimum distances from the z-axis as the width $d$ of the ergospheres, it can be observed from the right of the middle panel that the width of the ergospheres of the first excited state Dirac stars increases as the frequency decreases. In particular, for the excited state with $m_D=3/2$, the region of ergosphere existence is relatively short (frequency range $0.703\sim0.706$), and the ergosphere is nearly absent.
 
	On the other hand, from the left of the middle panel, it can be observed that the ergospheres are asymmetric concerning the equatorial plane (We use red dashed lines to represent the intersections of the equatorial plane and the $\rho\text{-}z$ plane.). However, as seen in the figure, this property is not very pronounced for $m_D=5/2$. We provide additional cross-sectional plots of the ergospheres of DSs for two other azimuthal harmonic indexes in the bottom panel. For $m_D=3/2$ and $m_D=7/2$, it is evident that the ergospheres are asymmetric with respect to the equatorial plane. For $m_D=5/2$, the ergospheres reach a maximum value of $z_{max}=0.60602$ and a minimum value of $z_{min}=-0.50489$ in the z-direction, which is also asymmetric.
 
	%%%%%%%%%%%%%%%%%%%%%%%%%%%%%%%%%%%%%%%%%%%%%%%%%%%%%%%%%%	
	\subsection{$D_1S_0$ state Dirac-boson stars}
	%%%%%%%%%%%%%%%%%%%%%%%%%%%%%%%%%%%%%%%%%%%%%%%%%%%%%%%%%%	
    	In this section, we introduce the ground state scalar field without nodes and investigate its influence on the first excited state Dirac stars, essentially the investigation of $D_1S_0$ state DBSs. Based on the relationship between the frequencies of the spinor field and the scalar field, we classify the DBSs into two categories: synchronized frequency solutions ($\omega_S=\omega_D$) and non-synchronized frequency solutions ($\omega_S \neq \omega_D$).
	%%%%%%%%%%%%%%%%%%%%%%%%%%%%%%%%%%%%%%%%%%%%%%%%%%%%%%%%%%	
	\subsubsection{Synchronized frequency}
	%%%%%%%%%%%%%%%%%%%%%%%%%%%%%%%%%%%%%%%%%%%%%%%%%%%%%%%%%%	
 	%
	\begin{figure}[!htbp]
		\centering
		\subfloat{  % []内可单独为每个小图命名。默认按照(a)(b)...的顺序命名，若省去[]则小图不命名。
			\includegraphics[width=7cm]{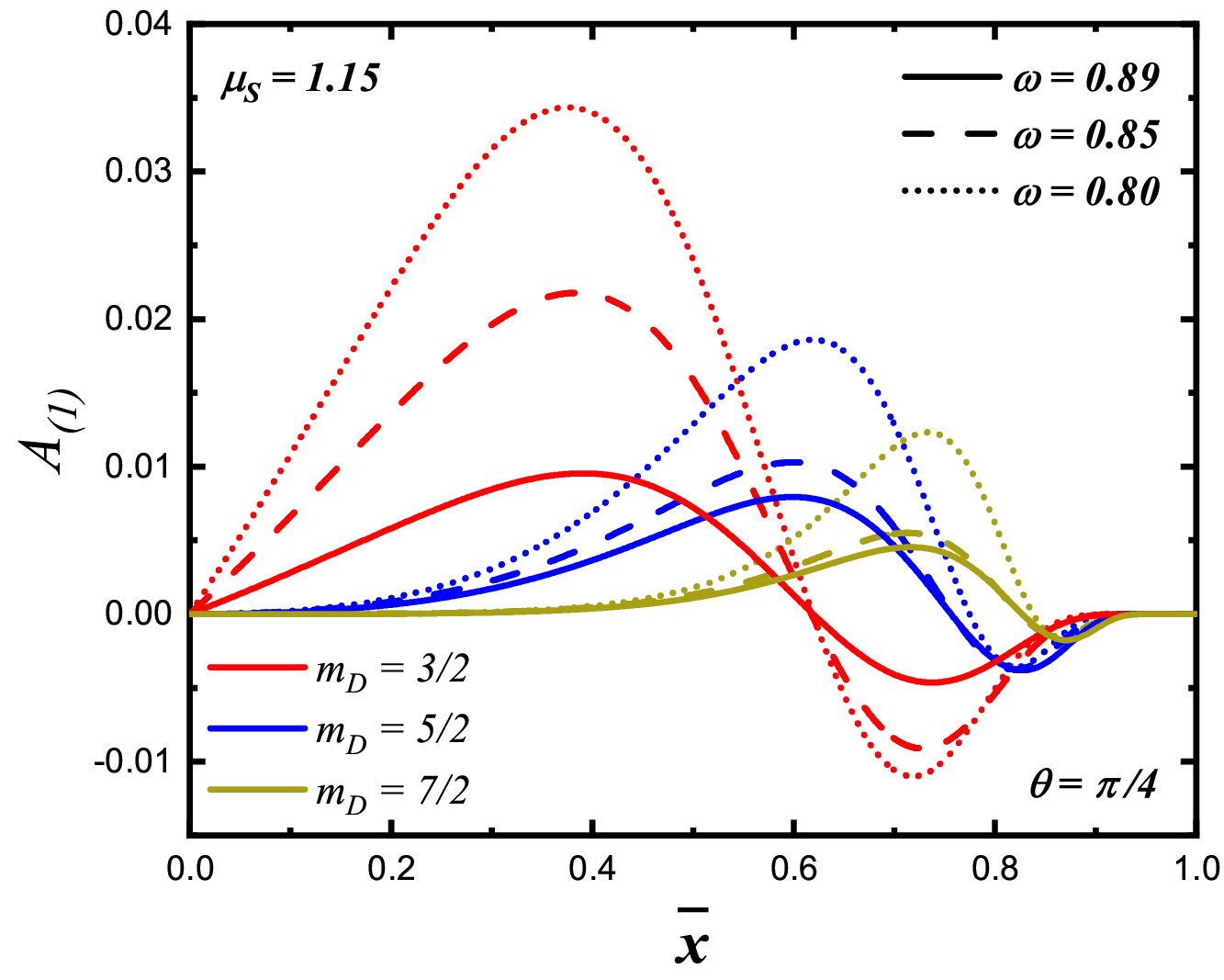}}
		\subfloat{
			\includegraphics[width=7cm]{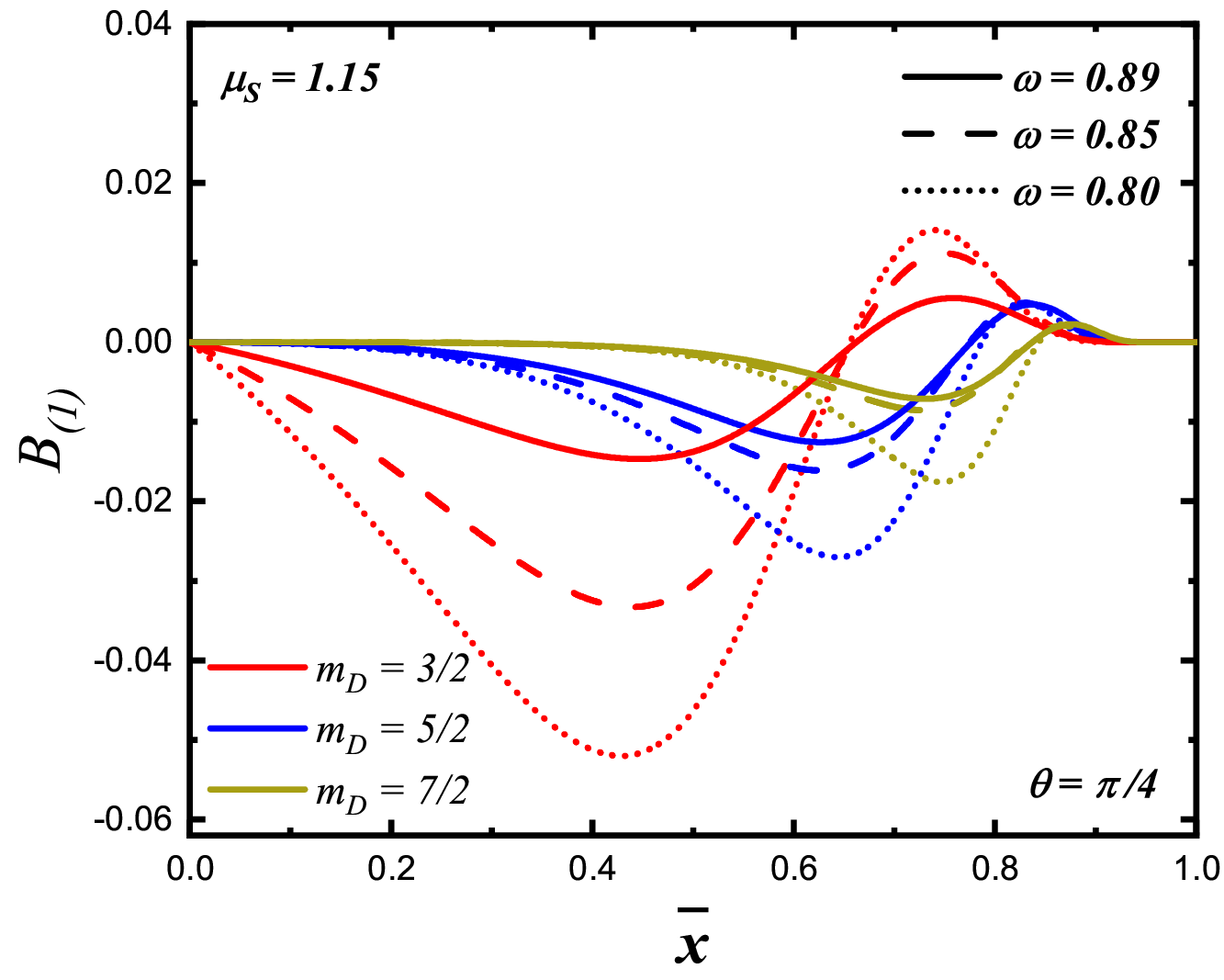}}
		\quad
		\subfloat{
			\includegraphics[width=7cm]{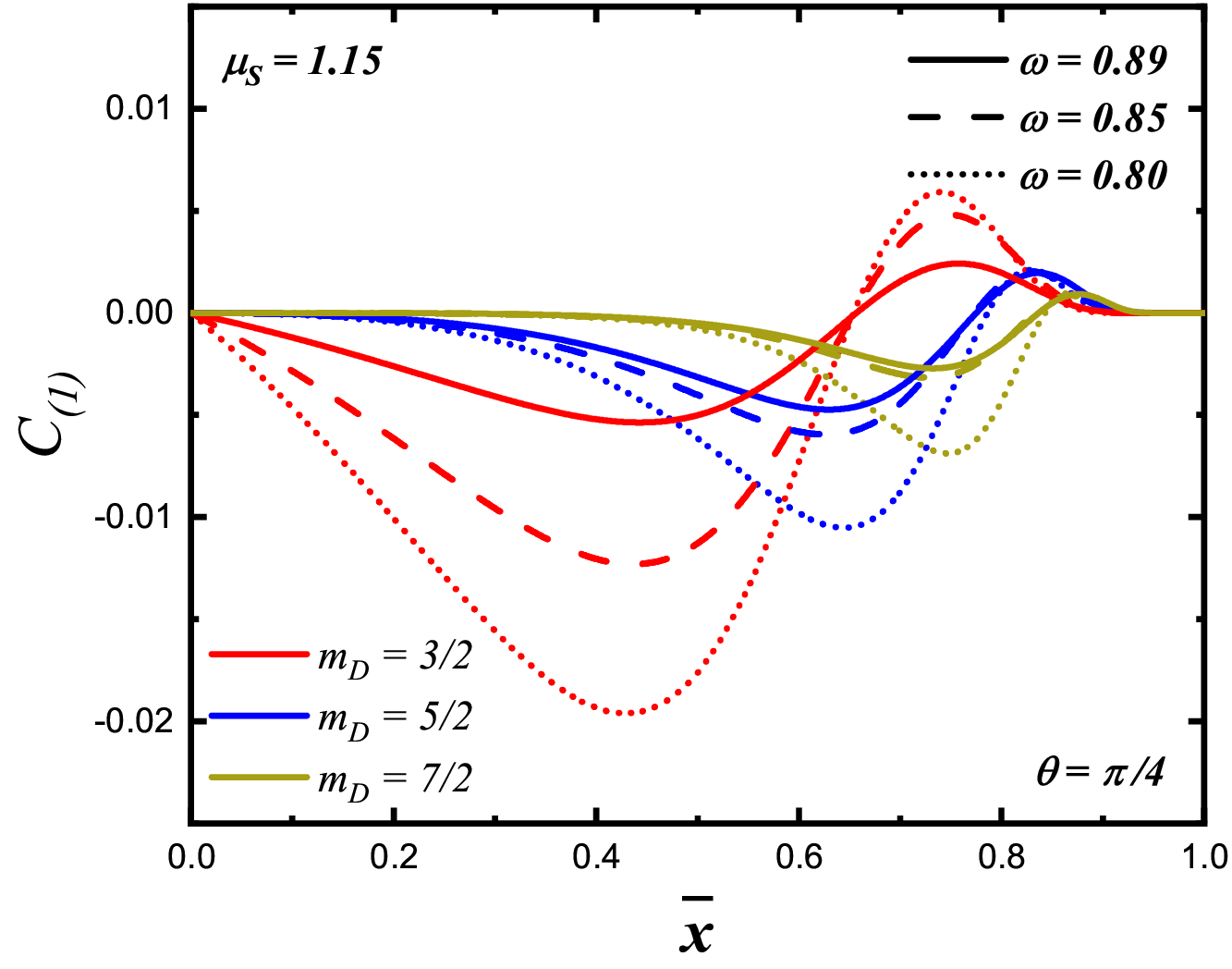}}
		\subfloat{
			\includegraphics[width=7cm]{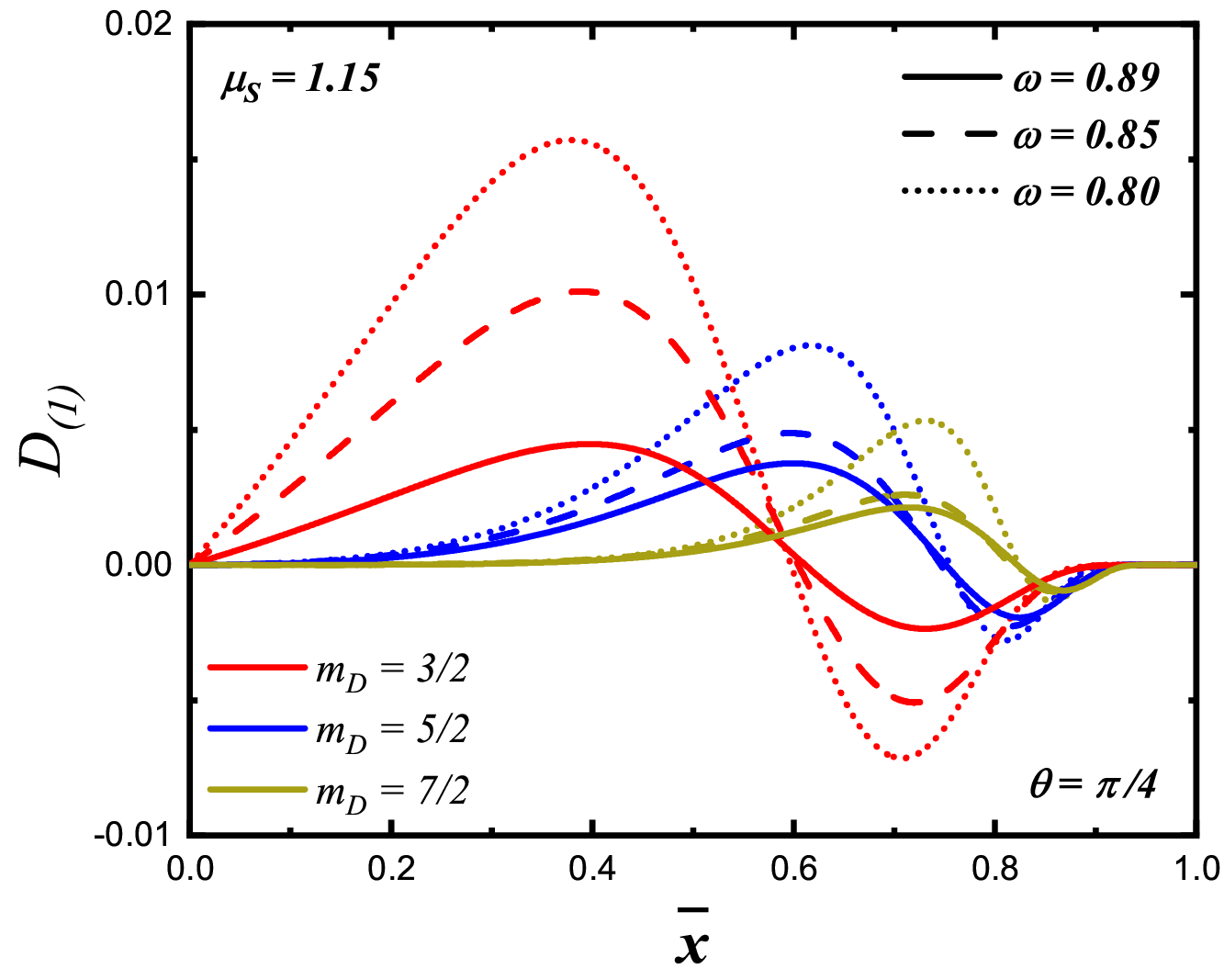}}
		\quad
		\subfloat{
			\includegraphics[width=7cm]{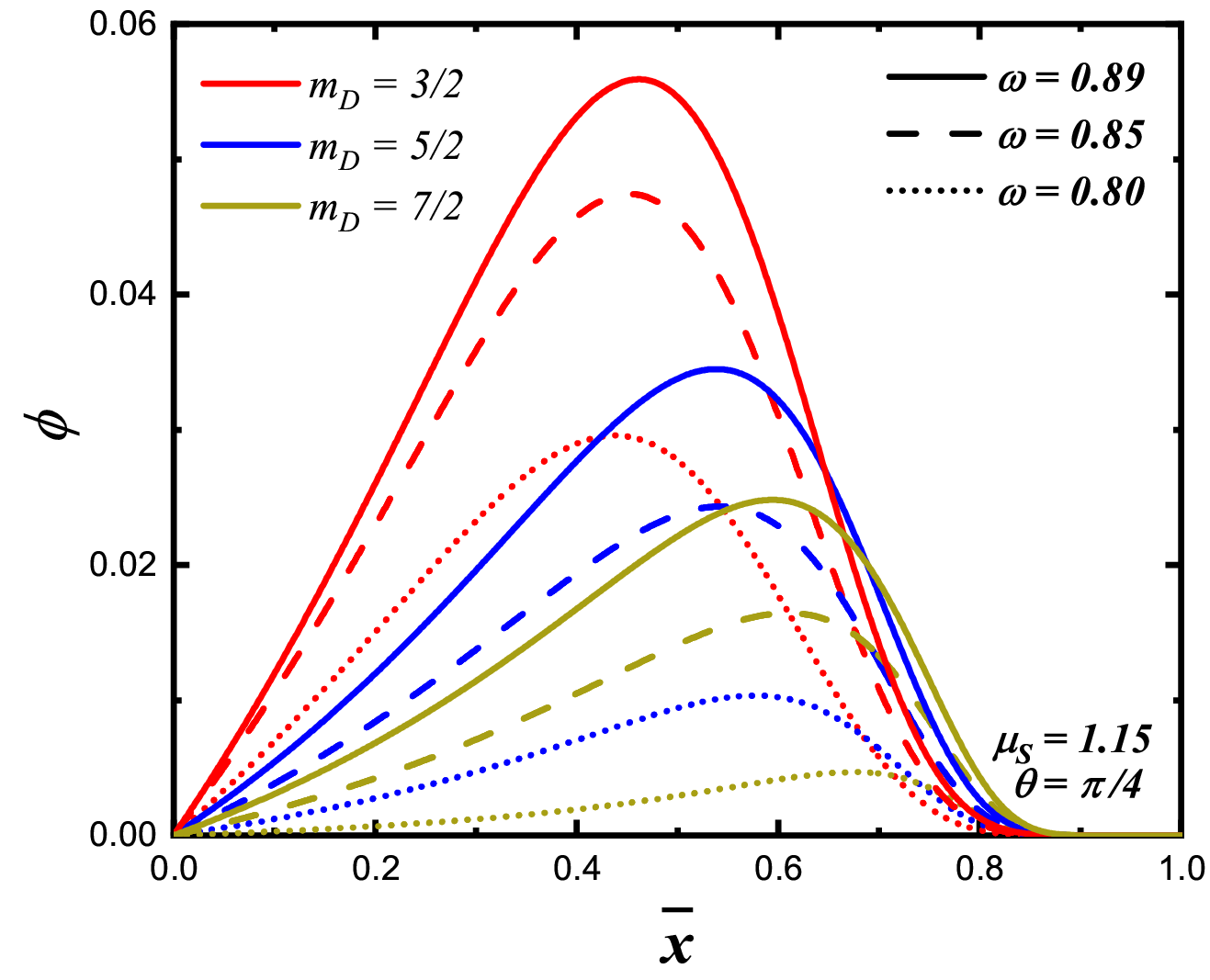}} 
		\subfloat{
			\includegraphics[width=6.5cm]{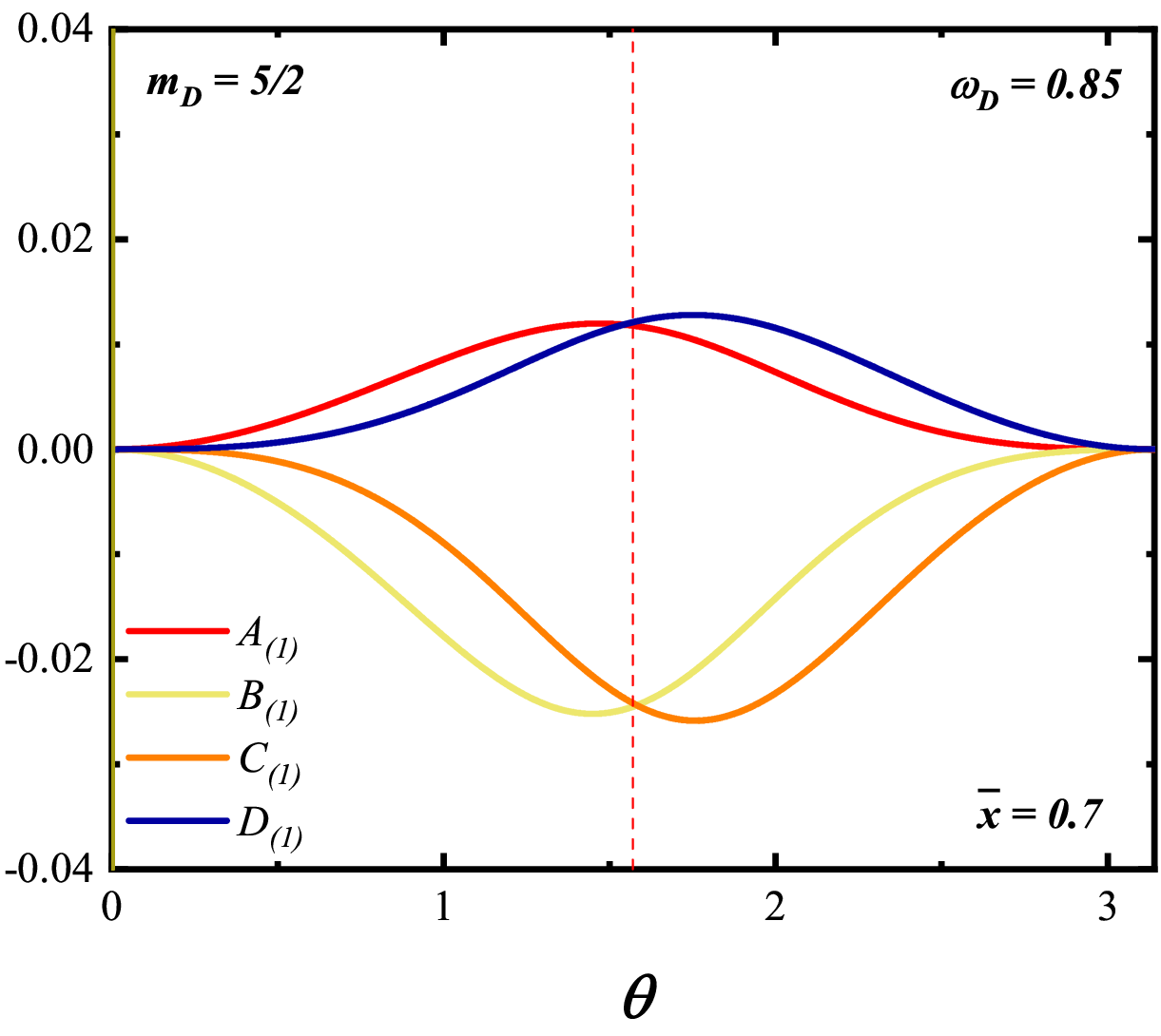}}  
		\caption{ The first excited spinor field functions $A_{(1)}$, $D_{(1)}$, $C_{(1)}$ and $D_{(1)}$ (Top and middle panels) and the ground scalar field function $\phi$ (Bottom left panel) of the DBSs as functions of $\bar{x}=0.7$ with the synchronized frequency $\omega = 0.80, 0.85, 0.89$ and $\theta=\pi/4$. The bottom right panel shows four spinor field functions of the DBSs as functions of $\theta$ with the synchronized frequency $\omega = 0.85$ and $\bar{x}=0.7$.}
		\label{fig:FunctionOfSys}
	\end{figure}
	In this subsection, we discuss the DBSs formed when the frequency of the spinor field is equal to that of the scalar field. Fig. \ref{fig:FunctionOfSys} shows the plots of several DBSs at different synchronized frequencies ($\omega=0.80$ as dot-dashed line, $\omega=0.85$ as dashed line, and $\omega=0.89$ as solid line) for $\mu_D=1$ first excited state of the spinor field functions ($A_{(1)}, B_{(1)}, C_{(1)}, D_{(1)}$) and the ground state of the scalar field function $\phi$. All solutions have $\mu_S=1.15$, and the red, blue, and yellow lines correspond to $m_D=3/2$, $5/2$, $7/2$, respectively. As the synchronized frequency increases, maximum absolute values of the spinor field functions $(A_{(1)})^{max}_{abs}$, $(B_{(1)})^{max}_{abs}$, $(C_{(1)})^{max}_{abs}$, and $(D_{(1)})^{max}_{abs}$ continuously decrease, while $\phi^{max}_{abs}$ (maximum absolute value of the scalar field function) increases. Additionally, similar to the case of DSs, the maximum absolute values of the four spinor field functions (top and middle left panels) and the scalar field function (bottom panel) decrease as the azimuthal harmonic indices increase. In the last picture of Fig. \ref{fig:FunctionOfSys}, we present the plot of the spinor field function on $\bar{x}=0.7$ plane. It can be observed that, with the introduction of the scalar field, similar to the case of DSs, the maximum absolute value of the spinor field function of DBSs decreases as the azimuthal harmonic indices increase. Additionally, the Dirac field function is also not symmetric about the equatorial plane. 
	
\begin{figure}[!t]
		\centering
		\subfloat{  % []内可单独为每个小图命名。默认按照(a)(b)...的顺序命名，若省去[]则小图不命名。
			\includegraphics[height=6cm]{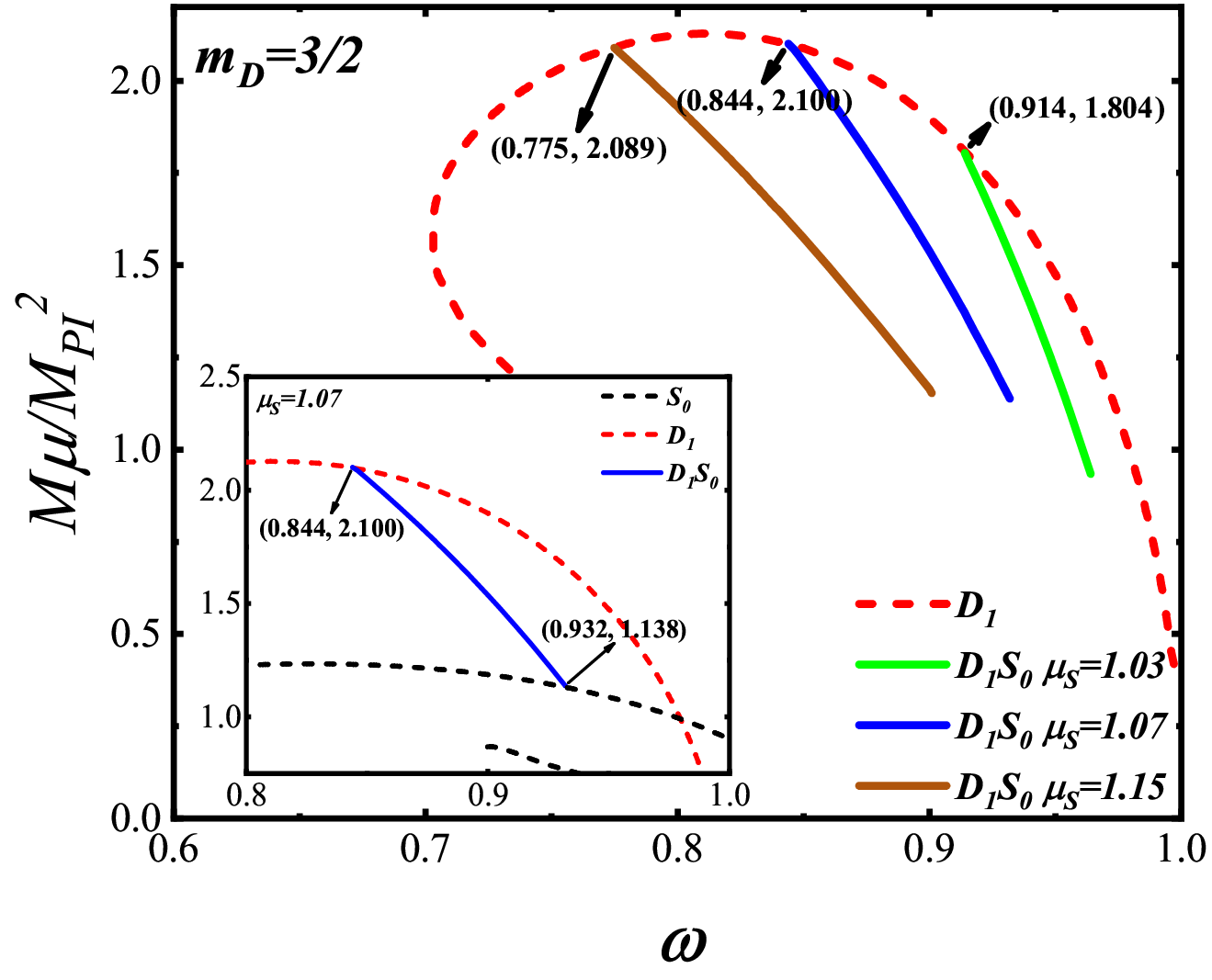}
		\label{fig:sys_a}}
		\subfloat{
			\includegraphics[height=6cm]{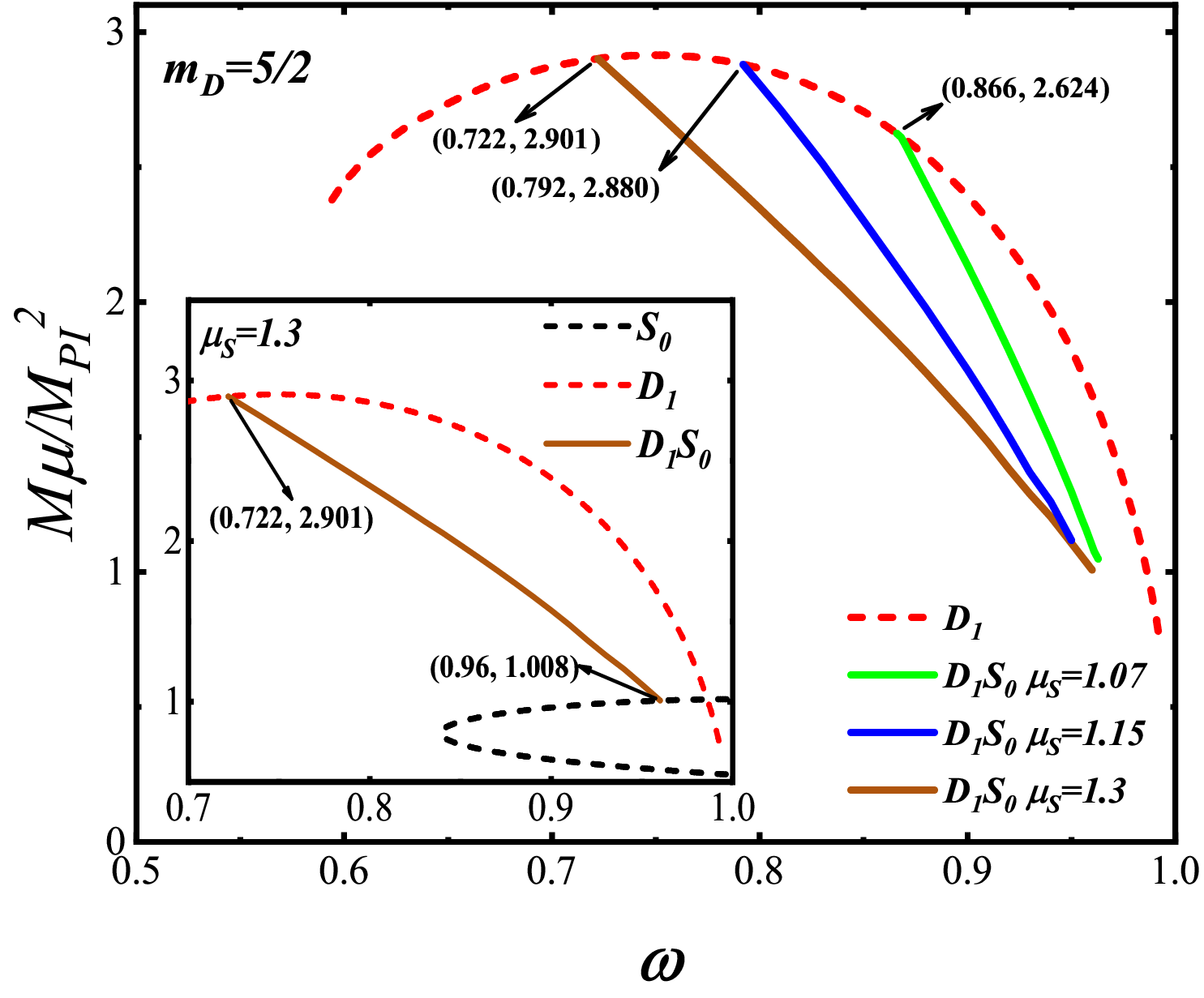}
		\label{fig:sys_b}}
        \quad
		\subfloat{
			\includegraphics[height=6cm]{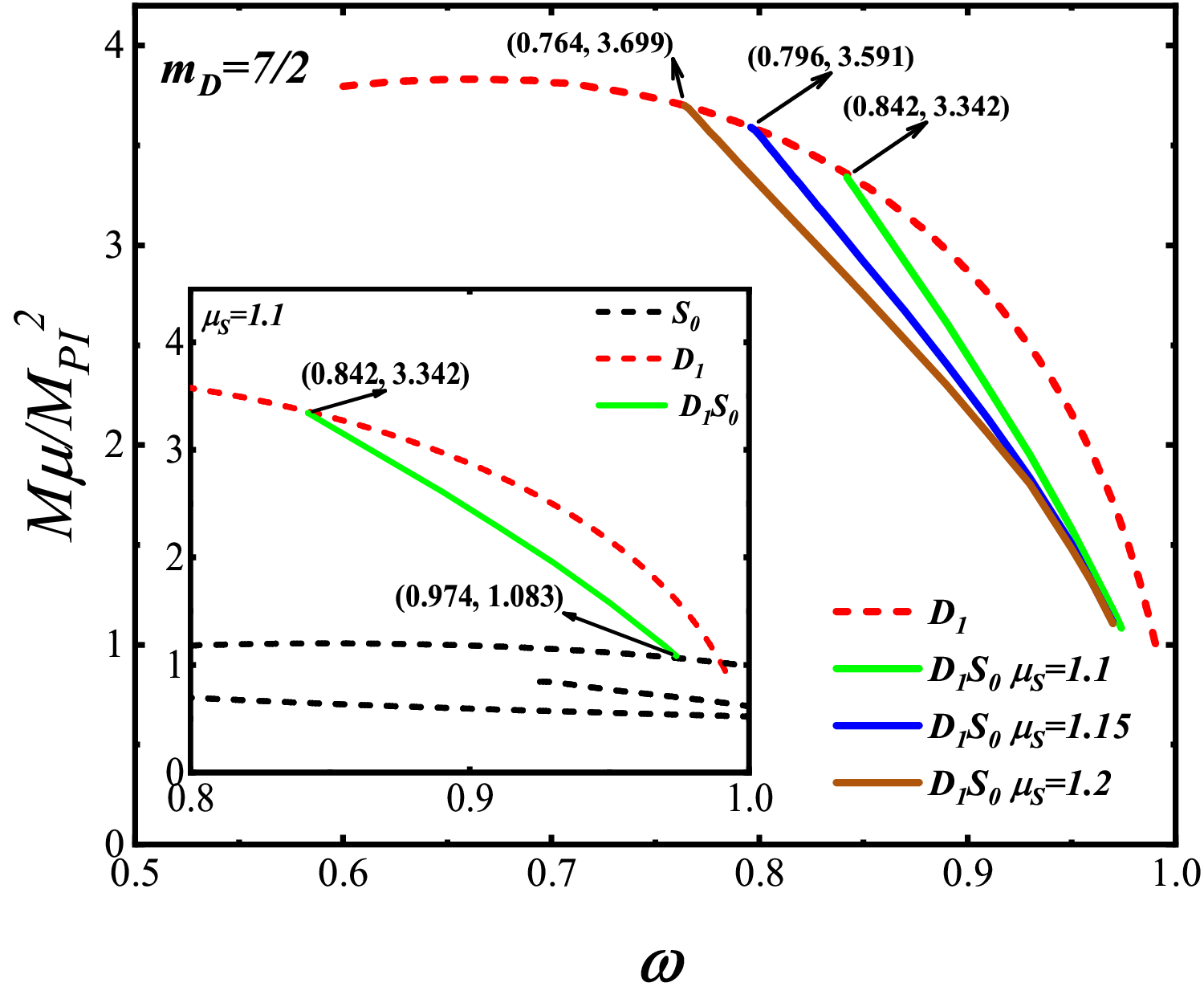}
		\label{fig:sys_c}}
		\caption{ The ADM mass $M$ vs. synchronized frequency $\omega$ of different azimuthal harmonic indexes DBSs. The red dashed line represents the first excited state of the Dirac stars (denoted by $D_1$). In the subfigure, the black dashed line represents the ground state of the boson stars (denoted by $S_0$), and the brown, blue, and green solid lines represent the multi-state $D_1S_0$ with $m_D=3/2$, $5/2$, $7/2$, respectively. For $m_D=3/2$ (top left panel), $\mu_S=1.03,1.07,1.15$; for $m_D=5/2$ (top right panel), $\mu_S=1.07,1.15,1.3$ , and for $m_D=7/2$ (bottom panel), $\mu_S=1.1,1.15,1.2$. All solutions have $\mu_D=1$.}
		\label{fig:sys}
	\end{figure}

	To further discuss the properties of DBSs, Fig. \ref{fig:sys} displays the relationship between the ADM mass $M$ and the synchronized frequency for multiple sets of DBSs with different azimuthal harmonic indices: $m_D=3/2$ (top left), $m_D=5/2$ (top right), and $m_D=7/2$ (bottom). The red lines represent the ADM mass of the excited state DBSs, while the black lines represent the mass of the ground state boson stars. Firstly, a second branch solution exists for the case of DSs with $m_D=3/2$. However, for the DBSs, we did not find any second branch solutions for $m_D=3/2$, $5/2$, $7/2$. Additionally, in the insets within the three panels, using a specific value of $\mu_S$ as an example, we depict the relationship curves between the ADM mass of DBSs and $\omega_D$, along with the intersection points with the first excited state DSs and the ground state boson stars. The coordinates of the intersections are as follows: for $m_D=3/2$ and $\mu=1.07$, they are located at ($0.844, 2.100$) and ($0.932, 1.138$); for $m_D=5/2$ and $\mu=1.15$, they are located at ($0.722, 2.901$) and ($0.960, 1.008$); for $m_D=7/2$ and $\mu_S=1.1$, they are located at ($0.842, 3.342$) and ($0.974, 1.083$). This implies that when the frequency is minor, the scalar field disappears, and the DBSs transition into Dirac stars. In Table \ref{tab: Sys}, we provide the existence domain of DBSs with respect to synchronous frequency for different scalar field masses, corresponding to three azimuthal harmonic indices. It can be observed that the existence domain of DBSs concerning synchronous frequency is smaller than that of DSs, and the mass reaches its maximum value when the frequency is minor. Finally, through an analysis of the ADM mass of DBSs as a function of $\omega_D$ for different values of the scalar field mass $\mu_S$, we can observe that the first excited state DBSs exhibit similar characteristics to those discussed in Ref. \cite{Liang:2022mjo} for the ground state DBSs. With an increase in the scalar field mass $\mu_S$, the minimum frequency value of DBSs decreases, indicating that the curve of ADM mass with respect to $\omega_D$ shifts to the left in the graph.

	\begin{table}[!t] 
		\centering 
		\begin{tabular}{|c|ccc|}

                \hline
			\multicolumn{4}{|c|}{$m_D=3/2$}   \\ \hline 
			& $\mu_S=1.03$ & $\mu_S=1.07$ & $\mu_S=1.15$ \\\hline
			$\omega$ & $0.914\sim0.964$ & $0.844\sim0.932$ & $0.775\sim0.901$ \\
			$M$ & $1.804\sim0.934$ & $2.100\sim1.138$ & $2.089\sim1.153$ \\\hline  \hline
			\multicolumn{4}{|c|}{$m_D=5/2$}  \\\hline
			& $\mu_S=1.07$ & $\mu_S=1.15$ & $\mu_S=1.30$ \\\hline
			$\omega$ & $0.866\sim0.963$ & $0.792\sim0.950$ & $0.722\sim0.960$ \\
			$M$ & $2.624\sim1.047$ & $2.880\sim1.118$ & $2.901\sim1.008$ \\\hline \hline
			\multicolumn{4}{|c|}{$m_D=7/2$}  \\\hline
			& $\mu_S=1.10$ & $\mu_S=1.15$ & $\mu_S=1.20$ \\\hline
			$\omega$ & $0.842\sim0.974$ & $0.796\sim0.970$ & $0.764\sim0.970$ \\
			$M$ & $3.342\sim1.083$ & $3.591\sim1.112$ & $3.699\sim1.105$\\\hline
		\end{tabular}
  		\caption{The domain of existence of the DBSs with the synchronized frequency $\omega_S=\omega_D=\omega$ in three different situations: the azimuthal harmonic indexes $m_D=3/2$, $5/2$, $7/2$. We set $\mu_S=1.03,1.07,1.15$ for $m_D=3/2$ (top panel), $\mu_S=1.07,1.15,1.30$ for $m_D=5/2$ (middle panel), and $\mu_S=1.10,1.15,1.20$ for $m_D=7/2$ (bottom panel) respectively. All solutions have $\mu_D=1$ and $m_S=1$. In all cases, the frequency decreases while the corresponding mass increases.}
        \label{tab: Sys}
	\end{table}

	%%%%%%%%%%%%%%%%%%%%%%%%%%%%%%%%%%%%%%%%%%%%%%%%%%%%%%%%%%	
	\subsubsection{Nonsynchronized frequency}
	%%%%%%%%%%%%%%%%%%%%%%%%%%%%%%%%%%%%%%%%%%%%%%%%%%%%%%%%%%	
	This section considers DBSs formed when the scalar field frequency does not change synchronously with the spinor field frequency. To analyze the influence of the parameters on the numerical solutions, we set the masses of the ground state and the excited state Dirac field to be the same ($\mu_S=\mu_D=1$). Due to our observation that multi-state DBSs transition into boson stars as the scalar field frequency exceeds $0.646$. At the same time, they do not degenerate into boson stars when the frequency is below $0.646$. Therefore, We classify these solutions into two categories based on this criterion: (1) degenerate solutions and (2) non-degenerate solutions.	
	%%%%%%%%%%%%%%%%%%%%%%%%%%%%%%%%%%%%%%%%%%%%%%%%%%%%%%%%%%	
     %
		    \begin{figure}[!htbp]
		\centering
		\subfloat{
			\includegraphics[width=7cm]{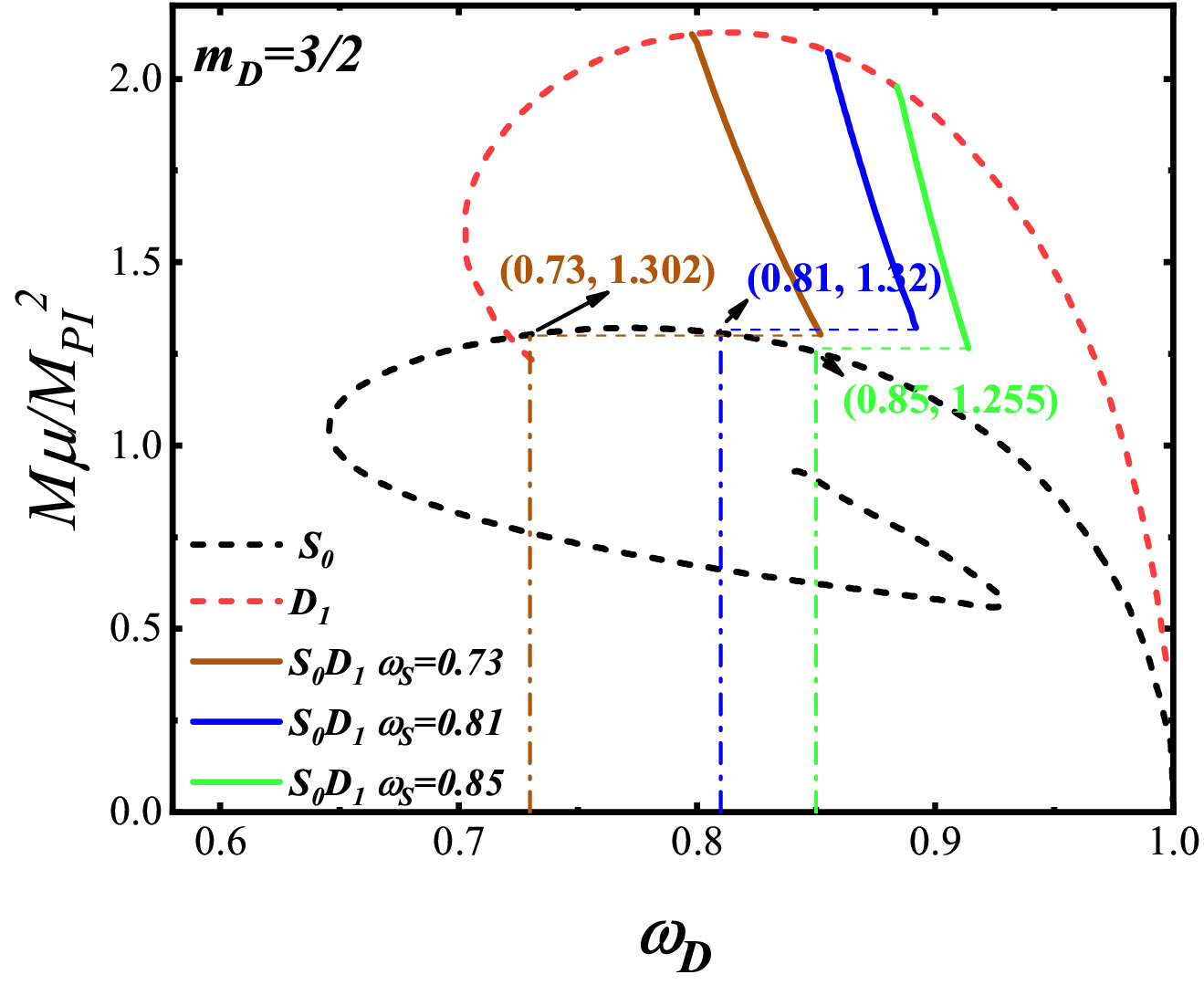}
			%\caption{fig1}
		}
		\subfloat{
			\includegraphics[width=7cm]{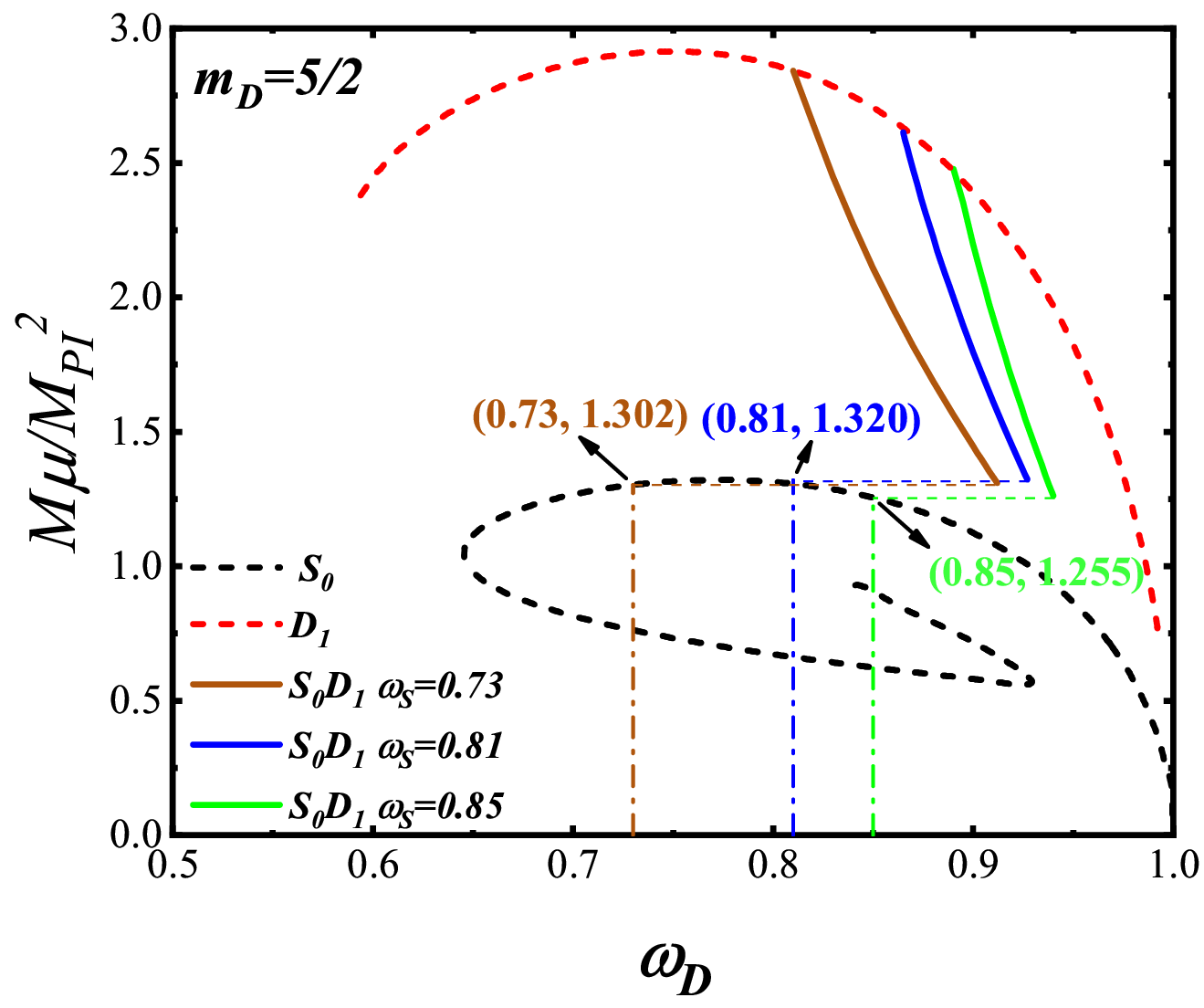}
		}
		\quad
		\subfloat{
			\includegraphics[width=7cm]{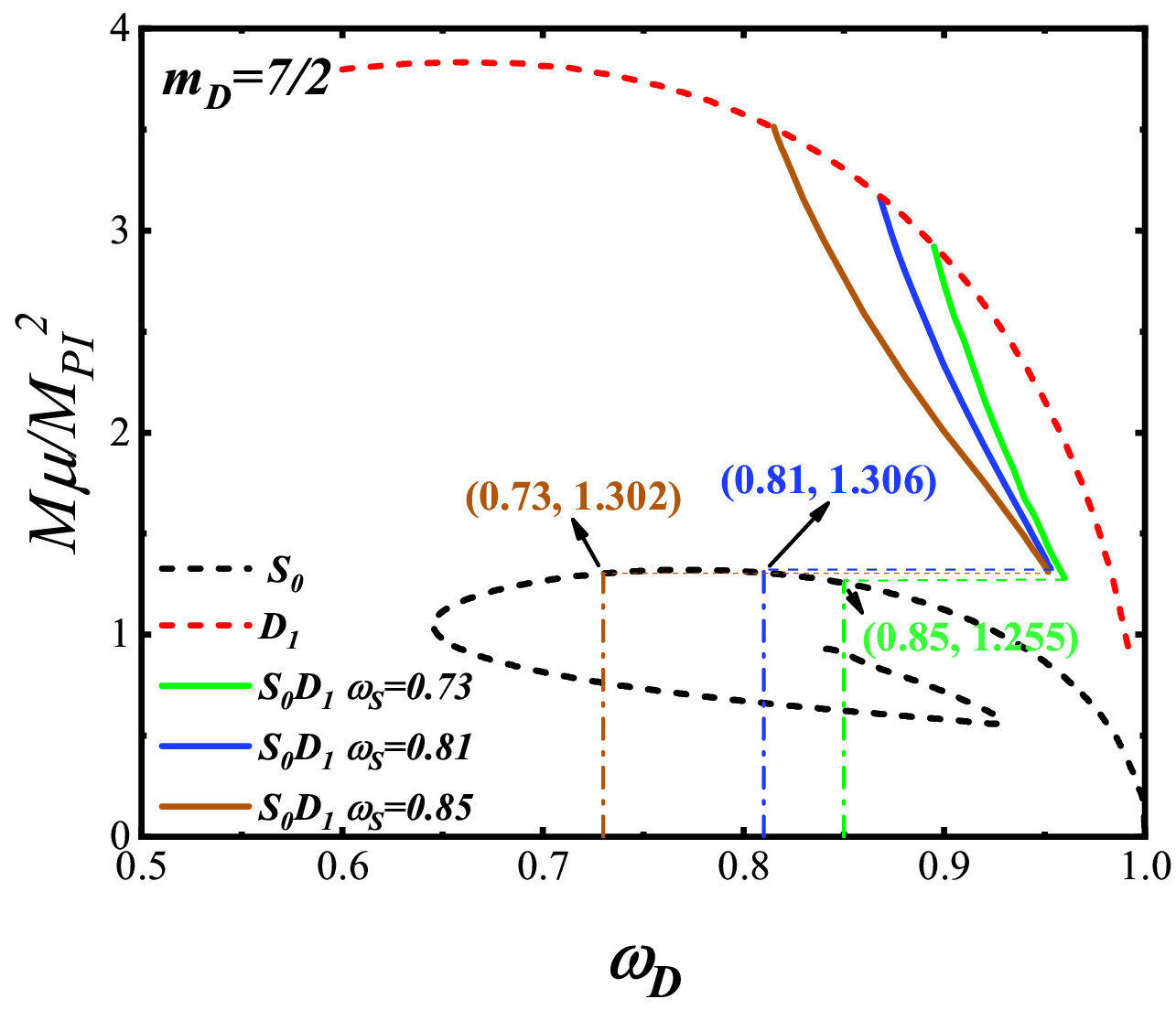}
		}
		\caption{ For the case of degenerate solutions with a higher scalar field frequency, the ADM mass $M$ of DBSs as a function of the nonsynchronized frequency $\omega_D$. The red dashed line represents the first excited state of the Dirac stars $D_1$, the black dashed line represents the ground state of the boson stars $S_0$, and the brown, blue, and green solid lines represent the multi-state $D_1S_0$ with $\omega_S=0.73,0.81,0.85$ respectively. All solutions have $\mu_D=\mu_S=1$ and $m_S=1$.}
		\label{fig:NonSyscan}
	\end{figure}

    In Fig. \ref{fig:NonSyscan}, we present the curves of ADM mass as a function of spinor field frequency for three different azimuthal harmonic indexes. In all cases, the red dashed line represents the excited state Dirac stars, the black dashed line represents the ground state boson stars (with its corresponding abscissa being its frequency $\omega_S$), and the solid lines in green ($\omega_S=0.73$), blue ($\omega_S=0.81$), and brown ($\omega_S=0.85$) represent the multi-state DBSs at three different frequencies. Similar to the case with synchronized frequencies, on the one hand, as the Dirac field frequency decreases, the mass of multi-state DBSs also increases, eventually evolving into excited-state Dirac stars, as indicated by the intersection of the red dashed line and the solid lines of three different colors. On the other hand, as the frequency increases, the mass of multi-state DBSs decreases until they become ground-state boson stars. We denote the point at which the Dirac field becomes zero and the corresponding scalar field frequency by the intersections of the two dashed lines that share the same color as the multi-state DBSs. The ADM mass degenerating into boson stars at the three frequencies are $1.302$ ($\omega_S=0.73$), $1.120$ ($\omega_S=0.81$), and $1.255$ ($\omega_S=0.85$), respectively. For the case of degenerate solutions, Table \ref{tab: Nonsys} provides the range of degenerate solutions of DBSs for the three azimuthal harmonic indexes at the three frequencies. 
    \begin{figure}[!t]
	\centering
	\subfloat{
		\includegraphics[width=7cm]{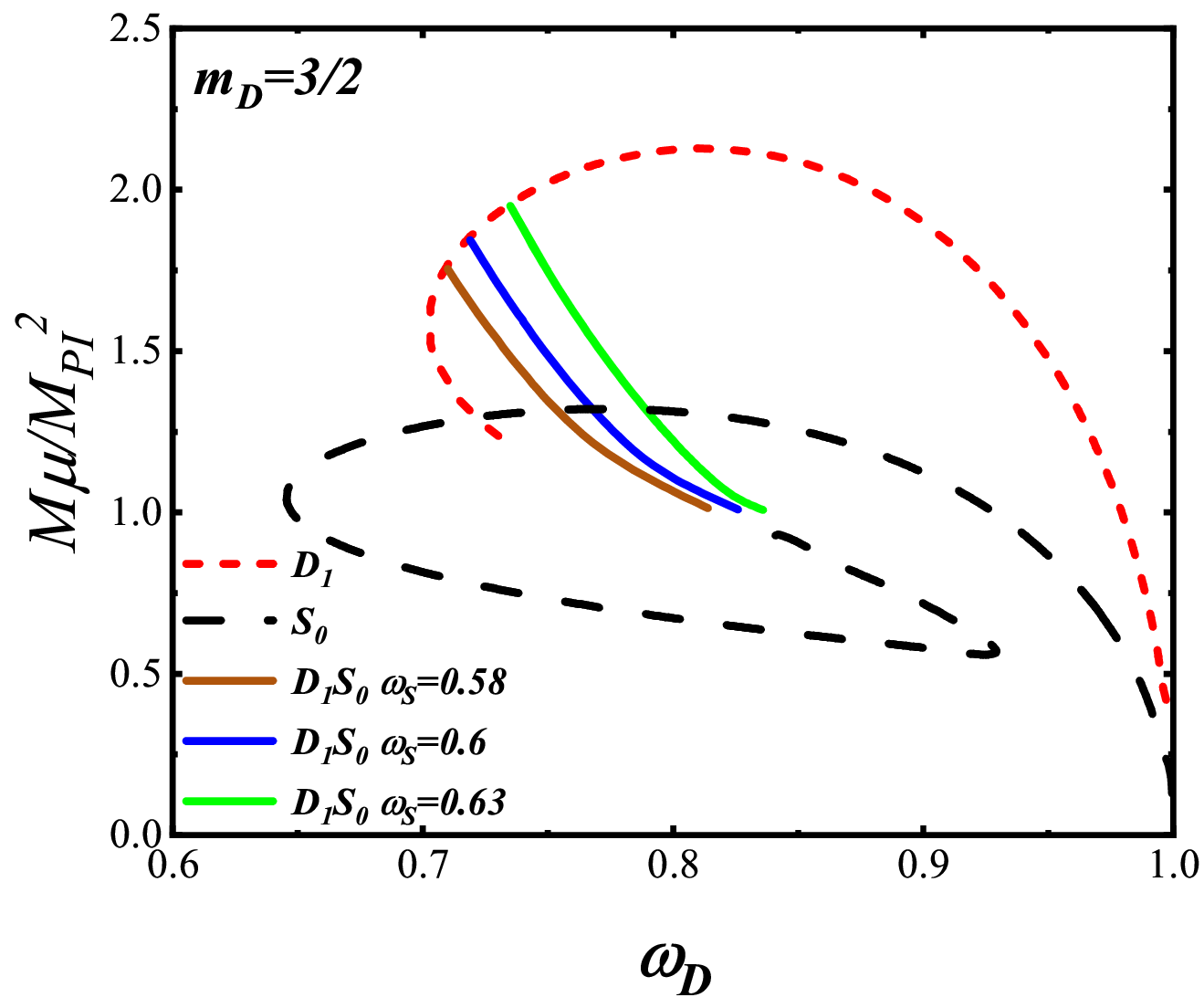}
		%\caption{fig1}
	}
	\subfloat{
		\includegraphics[width=7cm]{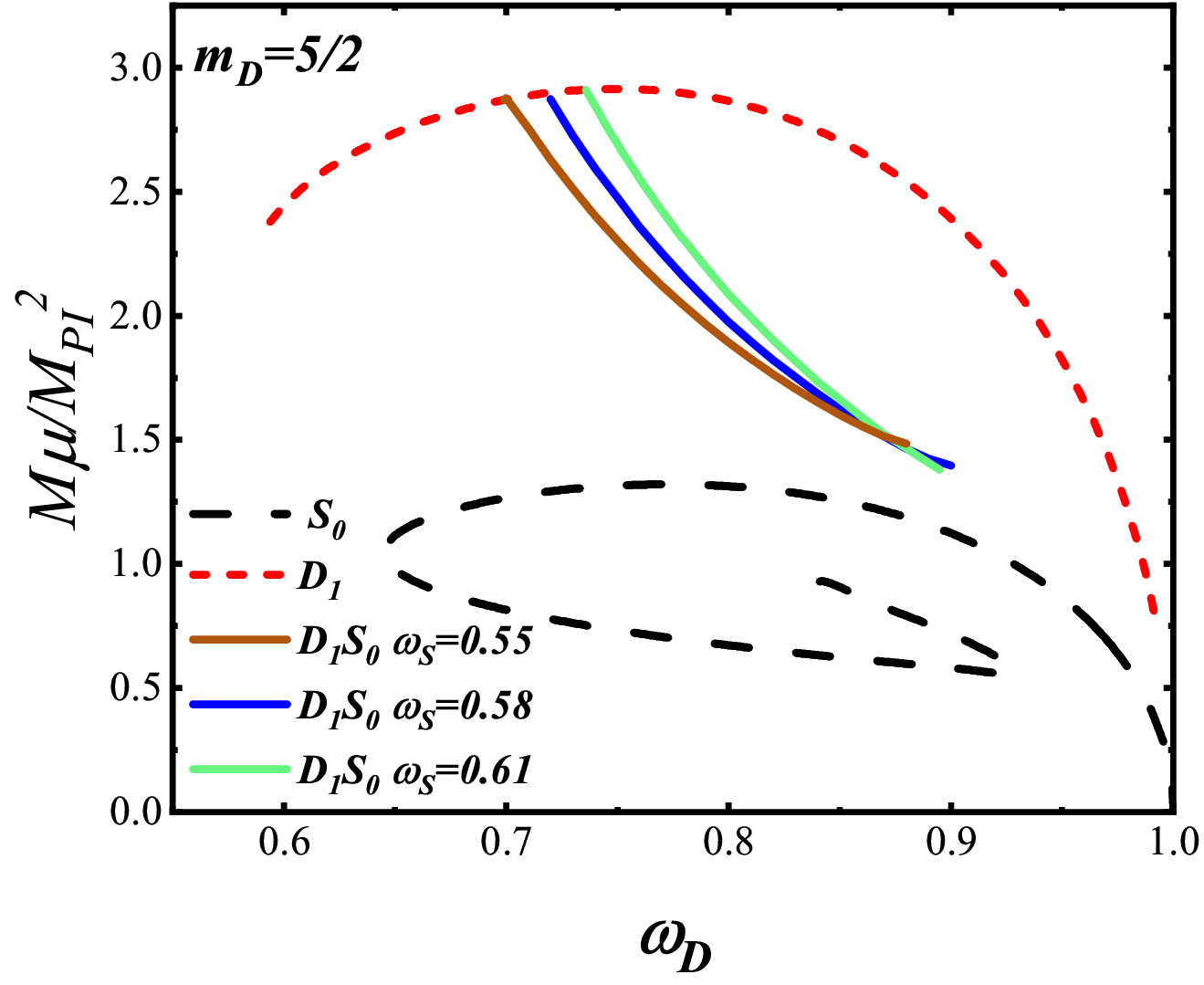}
	}
	\quad
	\subfloat{
		\includegraphics[width=7cm]{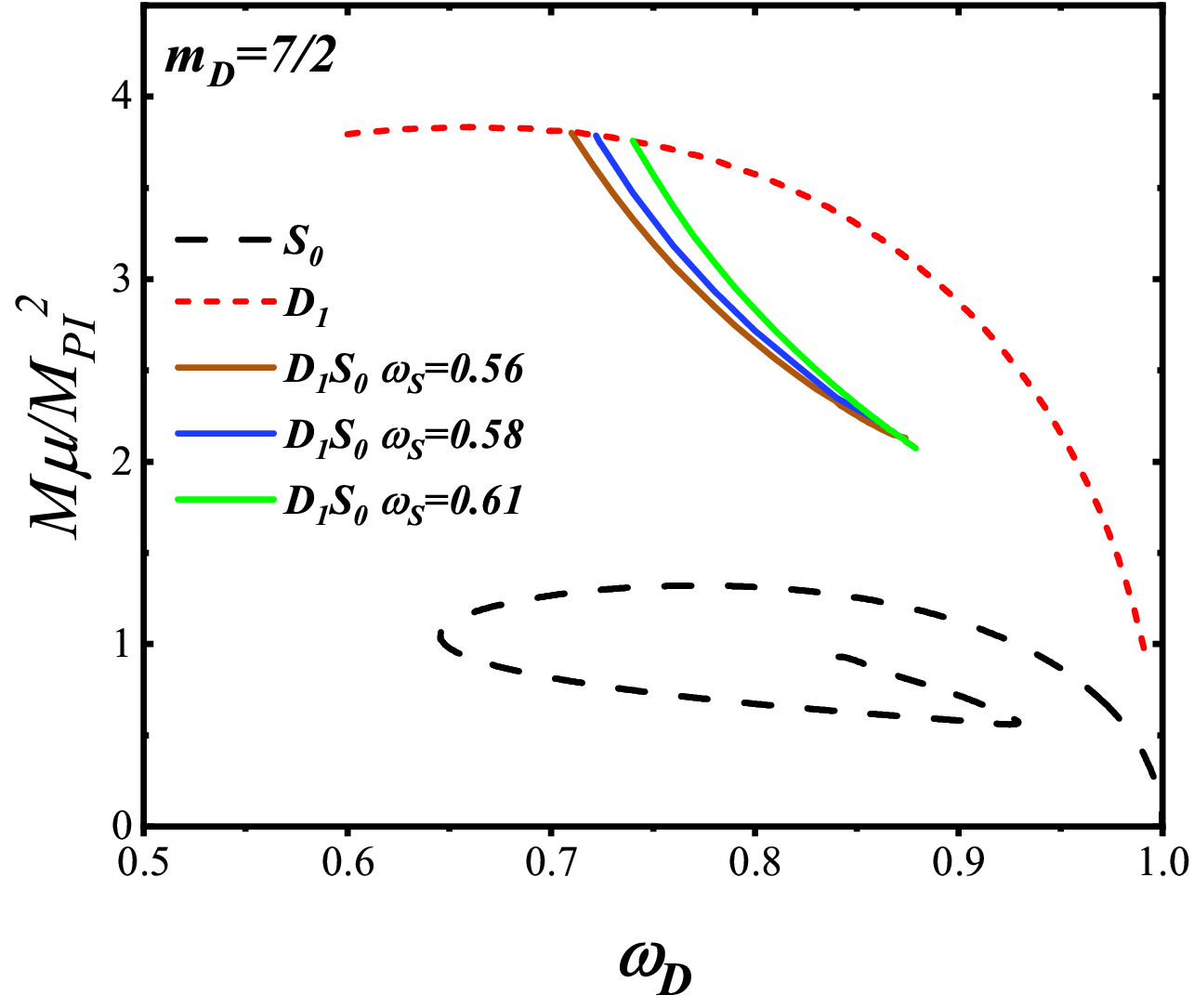}
	}
	\caption{  For the case of non-degenerate solutions with a lower scalar field frequency, the ADM mass $M$ of DBSs at the synchronized frequency $\omega_D$ is plotted. The red dashed line represents the first excited state of the spinor field $D_1$, the black dashed line represents the ground state of the scalar field $S_0$, and the brown, blue, and green solid lines represent the multi-state $D_1S_0$ with $\omega_S=0.67,0.7,0.73$ for $m_D=3/2$, $\omega_S=0.55,0.58,0.61$ for $m_D=5/2$, $\omega_S=0.56,0.58,0.61$ for $m_D=7/2$. All solutions have $\mu_D=\mu_S=1$ and $m_S=1$.}
	\label{NonSysDe2}
	\end{figure}
    %\small
    	\begin{table}[!htbp]     
    \scriptsize 
	\centering 
	\begin{tabular}{|c||cc|cc|cc|}
		\hline
		&\multicolumn{2}{c|}{$m_D=3/2$} &\multicolumn{2}{c|}{$m_D=5/2$} &\multicolumn{2}{c|}{$m_D=7/2$}     \\
		\hline
		& $\omega_D$ & $M$ & $\omega_D$ & $M$ & $\omega_D$ & $M$ \\
		$\omega_S=0.73$ & $0.798\sim0.852$ & $1.302\sim2.122$ & $0.810\sim0.912$ & $1.309\sim2.841$ & $0.815\sim0.952$ & $1.312\sim3.514$ \\
		$\omega_S=0.81$ & $0.855\sim0.892$ & $1.320\sim2.073$ & $0.865\sim0.927$ & $1.323\sim2.613$ & $0.868\sim0.953$ & $1.329\sim3.164$ \\
		$\omega_S=0.85$ & $0.844\sim0.914$ & $1.265\sim1.979$ & $0.890\sim0.940$ & $1.261\sim2.476$ & $0.895\sim0.960$ & $1.279\sim2.920$ \\
		\hline
	\end{tabular}
 	\caption{The domain of existence of the of degenerate DBSs in three different situations: The nonsynchronized frequency  $\omega_S=0.73$ (left panel), the nonsynchronized frequency $\omega_S=0.81$ (middle panel), and the nonsynchronized frequency $\omega_S=0.85$ (right panel) with the azimuthal harmonic indexes $m_D=3/2$, $5/2$, $7/2$, respectively.  All solutions have $\mu_D=\mu_S=1$ and $m_S=1$. In all cases, the frequency decreases while the corresponding mass increases.}
	\label{tab: Nonsys}
\end{table}

	Furthermore, from Fig. \ref{fig:NonSyscan}, it can also be observed that as the scalar field frequency decreases, the ADM mass curves of multi-state DBSs with respect to $\omega_D$ continuously shift to the left. Different types of solutions will emerge when the curves of ADM mass as a function of spinor field frequency shift to the left to a certain position as $\omega_D$ varies. We find that when the scalar field frequency is below approximately $0.646$, the multi-state DBSs will not transition into the ground-state boson stars as the frequency increases. This can be understood since the minimum value of the ground state scalar field frequency is around $0.646$\cite{Herdeiro:2015gia}, and when the spinor field becomes zero, the resulting solution, if it were a ground state boson star, would have nowhere to go. As an example, we present the curves of ADM mass versus $\omega_D$ of non-degenerate solutions corresponding to different azimuthal harmonic indices when the frequency is less than $0.646$ in Fig. \ref{NonSysDe2}. Similar to previous sections, the red and black dashed lines represent the excited state Dirac stars and the ground state boson stars, respectively. The solid lines of three different colors represent the multi-state DBSs at different scalar field frequencies. Similarly, the scalar field function can become zero when the spinor field frequency decreases. However, unlike the previous case, the spinor field does not disappear as the frequency increases. It should be noted that in the left panel of the figure, although the black dashed line appears to intersect with the solid line, the spinor field at the intersection point is not zero. In fact, it only appears to intersect. In Ref.\cite{Liang:2022mjo}, this type of non-degenerate solution was also discussed in the context of spherically symmetric DBSs, where, similar to DSs, these non-degenerate solutions exhibit multi-branch. However, in our paper, due to limitations in computational precision, as the frequency $\omega_D$ of non-degenerate solutions increases to a certain extent, errors rapidly increase, and we are uncertain whether multi-branch solutions for rotating DBSs exist under these conditions.

	    \begin{figure}[t!]
		\centering%图片居中
        \subfloat[]{
			\includegraphics[width=7cm]{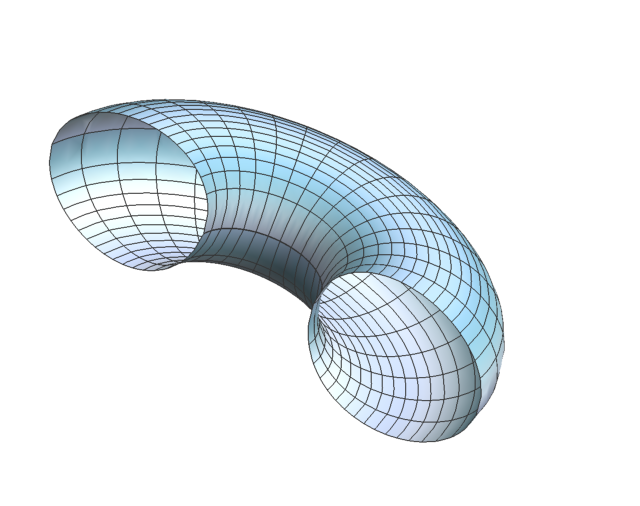}
      \label{fig: ergosphereNonsys_a}}  	
		\subfloat[]{
			\includegraphics[width=7cm]{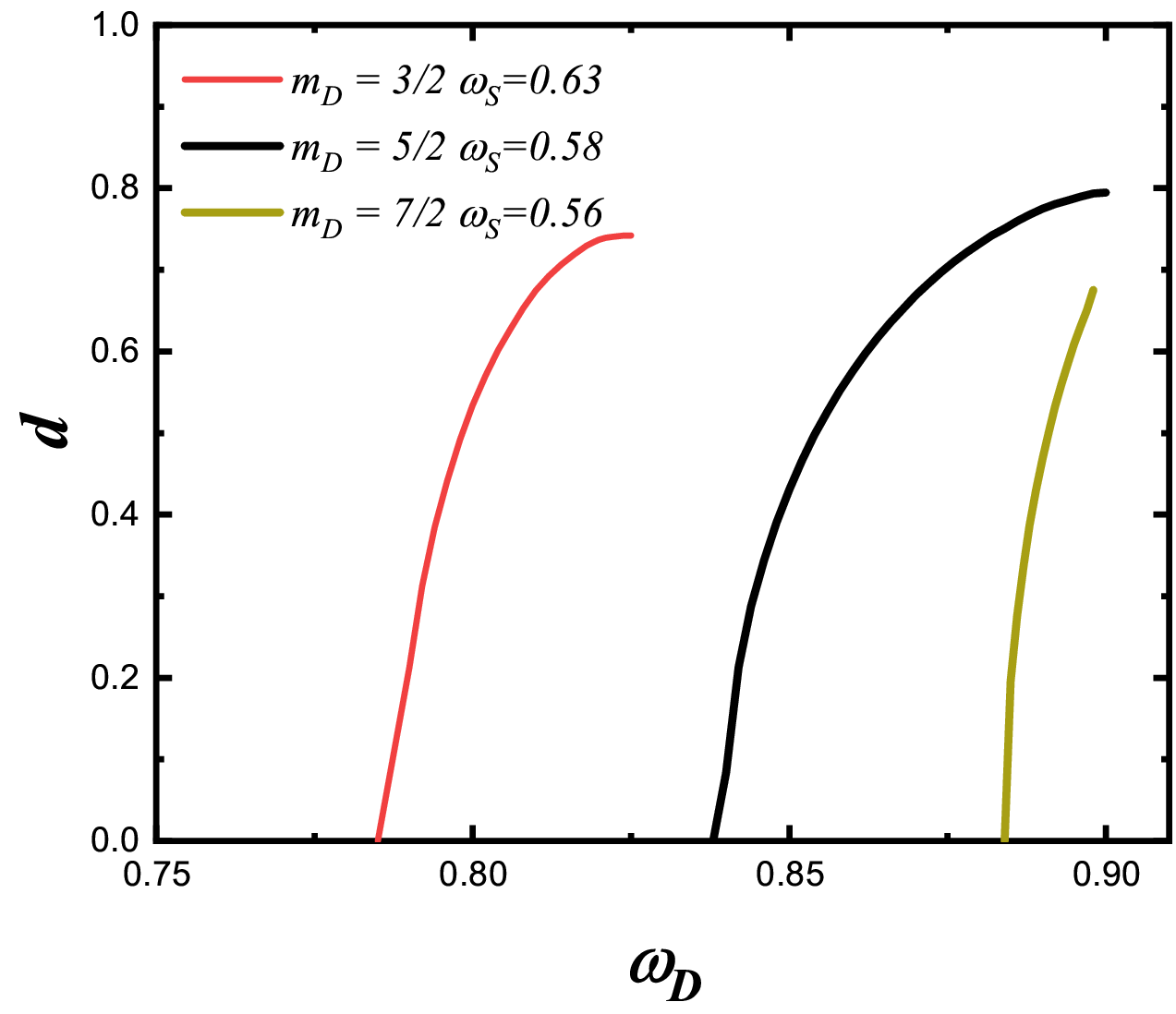}
      \label{fig: ergosphereNonsys_b}}
   \quad
        \subfloat[]{
			\includegraphics[width=7cm]{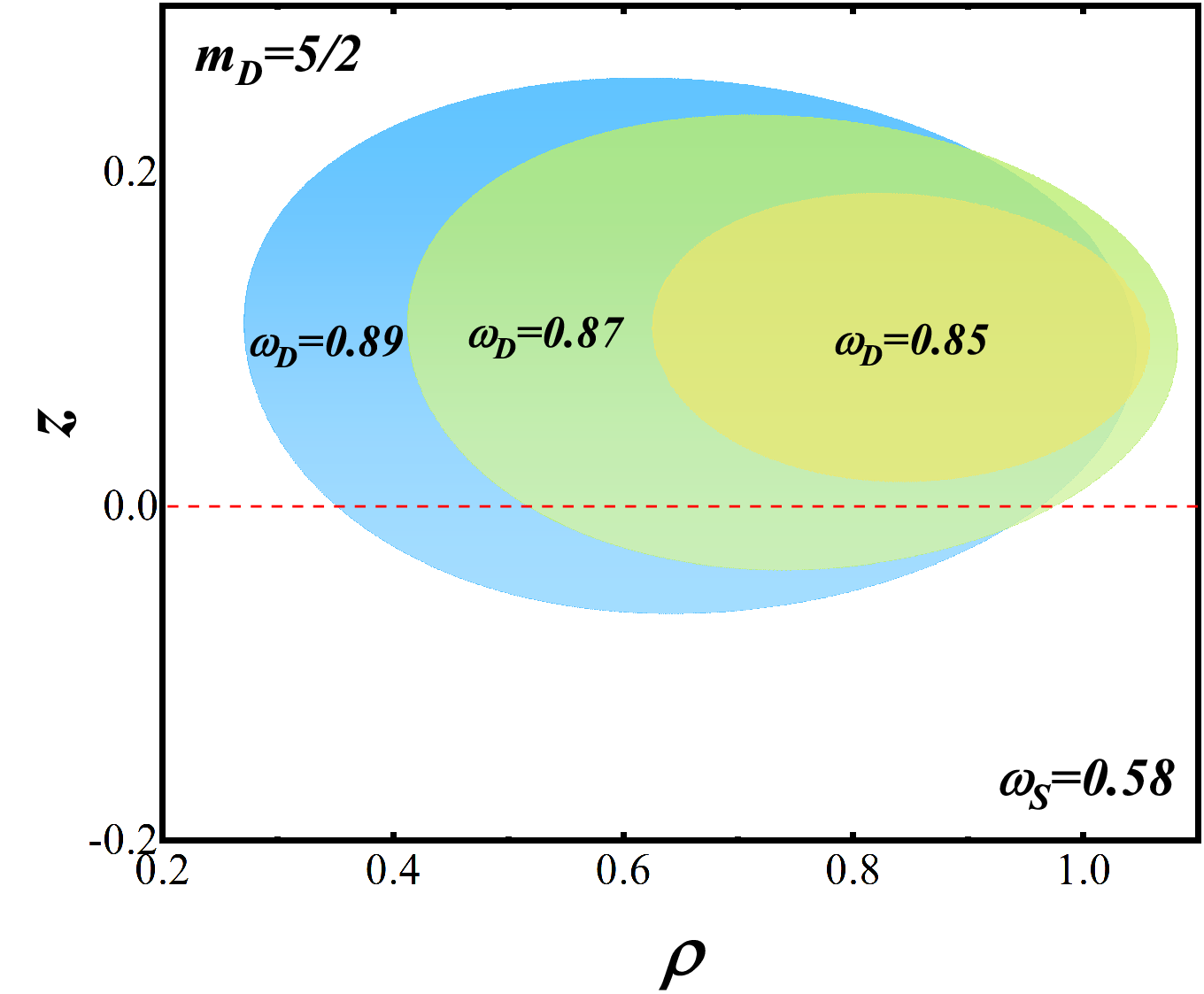}
      \label{fig: ergosphereNonsys_c}}
		\subfloat[]{
			\includegraphics[width=7cm]{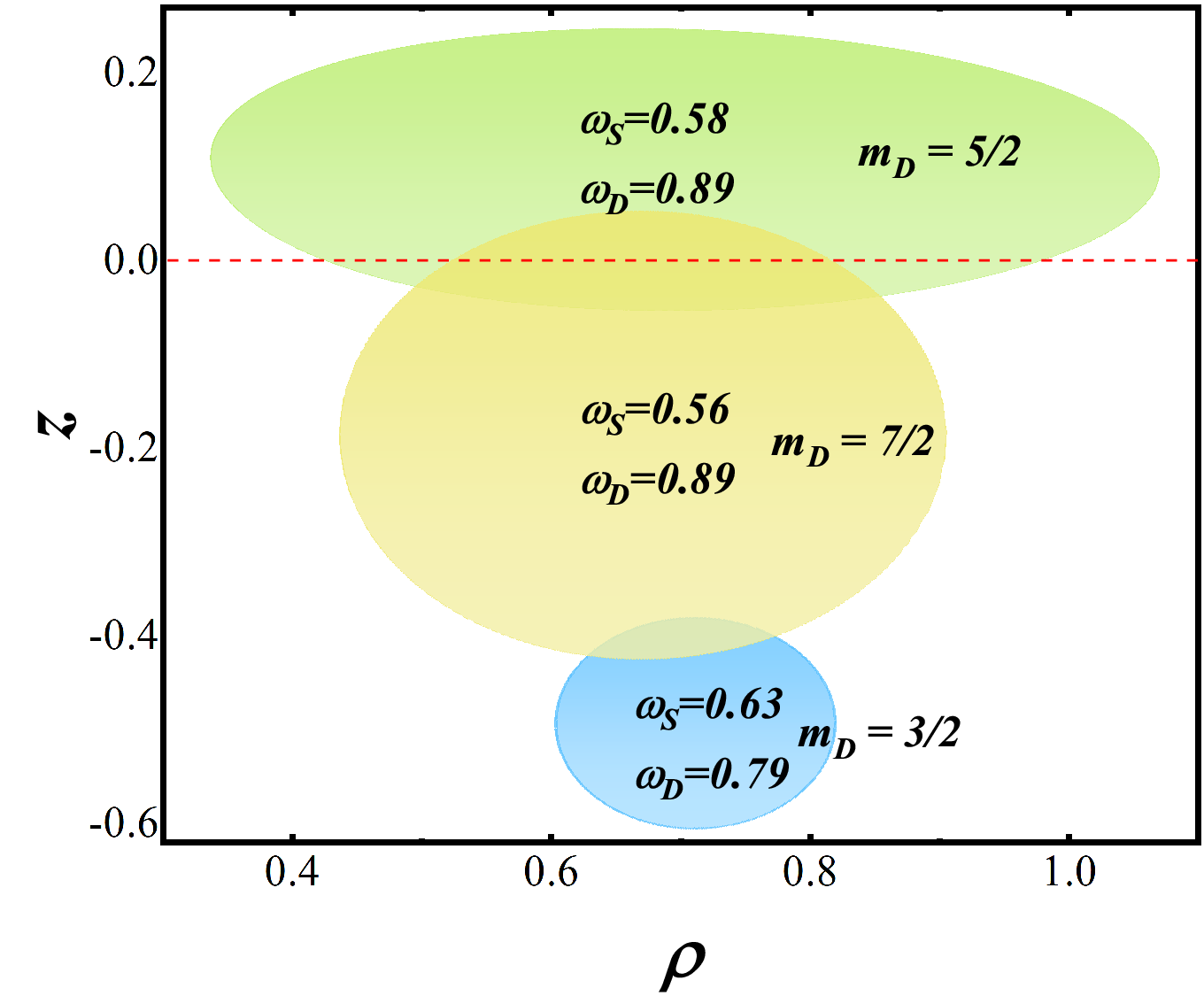}
   \label{fig: ergosphereNonsys_d}}
		\caption{Top left panel: Ergo-surface distribution in the domain of existence for the excited DBSs with $m_D = 5/2$ and $m_S=1$. Top right panel: The relationship between the width $d$ of the ergospheres and the frequency $\omega_D$ for DBSs with different azimuthal harmonic indexes. Bottom panel: Cross-sections of ergospheres for $m_D=5/2$ or on the $\rho\text{-}z$ plane with $\omega_D=0.85, 0.87, 0.89$. Bottom right panel: Cross-sections of the ergospheres for three different azimuthal harmonic indexes on the $\rho\text{-}z$ plane. }
		\label{fig: ergosphereNonsys}%与\ref连用，在文章与表格之间建立链接
	\end{figure}

	In the top left panel of Fig. \ref{fig: ergosphereNonsys}, we present the distribution of ergosurfaces for non-degenerate solutions. From the three-dimensional plot, we can observe that the ergospheres of DBSs exhibit a similar annular structure as the ergospheres of DSs. The bottom panel of Fig. \ref{fig: ergosphereNonsys} presents the cross-sectional of the DBSs' ergospheres on the $\rho\text{-}z$ plane ($\rho>0$).  The bottom left panel of Fig. \ref{fig: ergosphereNonsys} illustrates, using $m_D=5/2$ as an example, that as the frequency increases, the center of the ergospheres' cross-section continuously moves towards the left side of the graph (i.e., the $z$-axis), which is in opposite to the behavior of DSs. To further demonstrate the properties of the ergosurface, we provide in the bottom right figure of Fig. \ref{fig: ergosphereNonsys_d} the cross-sectional views of the ergospheres concerning the $\rho\text{-}z$ plane for $m_D=3/2, 5/2, 7/2$. It should be emphasized that due to the complexity of multi-state DBSs and the technical challenges in computing the ergospheres, achieving the same frequencies for all three azimuthal harmonic indices is difficult. From the bottom panel of Fig. \ref{fig: ergosphereNonsys}, it can also be observed that similar to the case of Dirac stars, the ergospheres of multi-states are also asymmetric with respect to the equatorial plane. However, unlike Dirac stars, Fig. \ref{fig: ergosphereNonsys_b}, for the three types of azimuthal harmonic indexes, we observe a difference in the behavior of the width $d$ of the ergospheres for DSs and DBSs as a function of frequency. For multi-states, the width $d$ increases with increasing frequency. This implies that, in general, as the solutions of the multi-state DBSs deviate from the solutions of individual Dirac stars, the width of the ergospheres becomes relatively large. It is worth noting that in this paper, only this type of non-degenerate DBSs exhibit ergospheres, while the two previous types of solutions (synchronized frequency solutions and degenerate solutions) do not possess ergosphere.

	%%%%%%%%%%%%%%%%%%%%%%%%%%%%%%%%%%%%%%%%%%%%%%%%%%%%%%%%%%
	\section{CONCLUSION}
	\label{sec6}		
	%%%%%%%%%%%%%%%%%%%%%%%%%%%%%%%%%%%%%%%%%%%%%%%%%%%%%%%%%%	
	In this paper, we first obtained the solutions of the rotationally symmetric first-excited state Dirac stars for higher azimuthal harmonic indices. We studied their physical parameters and the ergospheres. To study the influence of other matter fields on it, we introduce a scalar field and construct Dirac-boson stars composed of the spinor field of the first excited state and the scalar field of the ground state. 
	
	For the first excited state DSs, we find that their four field functions are asymmetric with respect to the equatorial plane, and the maximum values of the absolute values of the field functions decrease with increasing azimuthal harmonic index. On the one hand, by analyzing the curves of their ADM mass, angular momentum, and particle number as a function of frequency, we observe that the curves start from the vacuum solutions and gradually approach the minimum frequency. Specifically, for the case of $m_D = 3/2$, we also observed that after reaching the minimum frequency, the frequency increases again to a secondary maximum frequency. On the other hand, if the Pauli exclusion principle is enforced, which imposes single-particle conditions, it can be observed that the mass of the spinor field $\mu_D$ and the ADM mass $M$ will be finite, and they increase with the increase of the azimuthal harmonic indexes. In addition, the ergospheres of the first excited state DSs are asymmetric about the equatorial plane, and the width $d$ of the ergospheres increases with the frequency decreases.
	
	Furthermore, after introducing the scalar field, we found in our calculations that the scalar field can only affect DSs above a certain frequency. DBSs do not exist below this frequency, so the frequency existence domain of DBSs is smaller than that of DSs. Secondly, through the analysis of the ergospheres of DSs and DBSs, we found that the ergospheres of multi-state DBSs appear at higher frequencies. As the frequency of the Dirac field increases, the ergospheres' width $d$ of DBSs increases, and the center of the ergospheres' cross-section moves towards the z-axis. However, for the first excited state DSs, the ergospheres appear at lower frequencies. As the frequency of the Dirac field increases, the ergospheres' width $d$ decreases, and the center of the ergospheres' cross-section moves away from the z-axis. Furthermore, after introducing the scalar field, the four field functions and ergospheres of DBSs are still asymmetric about the equatorial plane, and the maximum value of the absolute value of Dirac field functions will still decrease with increasing the azimuthal harmonic index.
	
	The charged Dirac stars will generate more intriguing phenomena than the neutral Dirac stars. The solutions for the ground state of the rotating charged Dirac star have been found \cite{Herdeiro:2021jgc}. Our next step will be to consider how the properties of the Dirac star will change if the excited state of the Dirac star is charged. Furthermore, Anti–de Sitter (AdS) spacetime is important in general relativity and modern field theory. Therefore, we also plan to consider the negative cosmological constant \cite{Kumar:2016oop, Giammatteo:2004wp} and investigate solutions of the Dirac-Einstein equations in AdS spacetime.

	%%%%%%%%%%%%%%%%%%%%%%%%%%%%%%%%%%%%%%%%%%%%%%%%%%%%%%%%%%
	\section*{ACKNOWLEDGEMENTS}
	%%%%%%%%%%%%%%%%%%%%%%%%%%%%%%%%%%%%%%%%%%%%%%%%%%%%%%%%%%
	This work is supported by National Key Research and Development Program of China (Grant No. 2020YFC2201503) and the National Natural Science Foundation of China (Grants No.~12275110 and No.~12047501). Parts of computations were performed on the Shared Memory system at Institute of Computational Physics and Complex Systems in Lanzhou University. 	
	
	%%%%%%%%%%%%%%%%%%%%%%%%%%%%%%%%%%%%%%%%%%%%%%%%%%%%%%%%%%%%%%%%%%
		\begin{small}

		%%%%%%%%%%%%%%%%%%%%%%%%%%%%%%%%%%%%%%%%%%%%%%%%%%%%%%%%%%%%%%%%%%%%%%%%%%%%%%%
		%%%%%%%%%%%%%%%%%%%%%%%%%%%%%%%%%%%%%%%%%%%%%%%%%%%%%%%%%%%%%%%%%%%%%%%%%%%%%%
	\end{small}
	
\end{document}